\documentclass[reprint,twocolumn,superscriptaddress,preprintnumbers,nofootinbib,aps,prd]{revtex4}

\usepackage{amsmath}
\usepackage{amssymb}
\usepackage{hyperref}
\usepackage{graphicx}

\usepackage{slashed}

\newcommand\td{\text{3d}}
\newcommand\fd{\text{4d}}

\begin{document}

\newcommand{\HIPetc}{\affiliation{
Department of Physics and Helsinki Institute of Physics,
PL 64,
FI-00014 University of Helsinki,
Finland
}}

\newcommand{\umassAff}{\affiliation{
Amherst Center for Fundamental Interactions, Department of Physics, University of Massachusetts, Amherst, MA 01003
}}

\title{Electroweak phase transition in the $\Sigma$SM - I: Dimensional reduction}

\preprint{ACFI-T18-04, HIP-2018-7/TH}

\author{Lauri Niemi}
\email{lauri.b.niemi@helsinki.fi}
\HIPetc

\author{Hiren H. Patel}
\email{hhpatel@umass.edu}
\umassAff

\author{Michael J. Ramsey-Musolf,}
\email{mjrm@physics.umass.edu}
\umassAff

\author{Tuomas V. I. Tenkanen}
\email{tuomas.tenkanen@helsinki.fi}
\HIPetc

\author{David J. Weir}
\email{david.weir@helsinki.fi}
\HIPetc

\date{\today}

\begin{abstract}
In a series of two papers, we make a comparative analysis of the performance of conventional perturbation theory to analyze electroweak phase transition in the real triplet extension of Standard Model ($\Sigma$SM).  In Part I (this paper), we derive and present the high-$T$ dimensionally reduced effective theory that is suitable for numerical simulation on the lattice.  In the sequel (Part II), we will present results of the numerical simulation and benchmark the performance of conventional perturbation theory.  Under the assumption that $\Sigma$ is heavy, the resulting effective theory takes the same form as that derived from the minimal standard model.  By recasting the existing non-perturbative results, we map out the phase diagram of the model in the plane of triplet mass $M_\Sigma$ and Higgs portal coupling $a_2$.  Contrary to conventional perturbation theory, we find regions of parameter space where the phase transition may be first order, second order, or crossover.  We comment on prospects for prospective future colliders to probe the region where the electroweak phase transition is first order by a precise measurement of the $h\rightarrow\gamma\gamma$ partial width.
\end{abstract}
\maketitle

\section{Introduction}
\label{sec:intro}

Explaining the origin of the observed baryon asymmetry of the universe, characterized by the  baryon to entropy density ratio,
\begin{equation*}
Y_B \equiv \rho_B / s = (8.61\pm0.09)\times 10^{-11} \enspace \text{\cite{Ade:2015xua}}
\end{equation*}
remains an outstanding problem at the interface of high energy and nuclear physics with cosmology.  General considerations identified by Sakharov \cite{Sakharov:1967dj} impose three criteria on early universe particle physics in order to explain the asymmetry: non-conservation of baryon number, violation of C and CP invariance, and presence of non-equilibrium conditions\footnote{The latter requirement assumes CPT invariance.}.  While the Standard Model (SM) of particle physics  supplies the baryon non-conserving interactions in the form of sphaleron processes, it provides neither the requisite non-equilibrium conditions nor sufficiently effective CP-violation. Thus, physics beyond the Standard Model (BSM) is essential.

Several mechanisms have been advanced that satisfy the required criteria.  Among the most compelling and theoretically well-motivated  is electroweak baryogenesis, wherein the baryon asymmetry is generated during the era of electroweak symmetry breaking (for a recent review, see Ref.~\cite{Morrissey:2012db}).  Successful baryogenesis requires that symmetry breaking occurred due a strongly first order electroweak phase transition.  Numerical lattice simulations \cite{Kajantie:1995dw,Gurtler:1997hr,Laine:1998jb,Csikor:1998eu,Rummukainen:1998as,Aoki:1999fi} indicate that EWSB in the SM occurred a through a crossover transition for a Higgs mass at its observed value of 126 GeV \cite{Chatrchyan:2012xdj, Aad:2012tfa}, suggesting that the universe never departed from thermal equilibrium during this epoch.

BSM scenarios may alleviate this SM shortcoming through the addition of an extended scalar sector. The latter may catalyze a strong first order electroweak phase transition (SFOEWPT) through new loop corrections to the zero temperature ($T$) Coleman-Weinberg potential, thermal loop corrections to the finite-$T$ effective potential, a modification of the tree level vacuum structure of the theory, or a combination involving more than one of these effects. The result may be not only a SFOEWPT to the present \lq\lq Higgs phase", but also a richer pattern of symmetry-breaking that precedes the Higgs phase than one obtains in the SM. 

These possibilities have been explored in both U.V. complete theories, such as the Minimal Supersymmetric Standard Model (MSSM), and simplified models that consider only the extended scalar sector. While simplified models are not realistic descriptions of nature, their use allows one to identify general features of phase transition dynamics that may occur in various U.V. complete theories and to delineate the corresponding phenomenological consequences. Perhaps, the most widely considered such simplified model involves the addition of a real scalar that carries no SM gauge charge. The phase transition dynamics of the singlet-extended Standard Model (xSM) and corresponding implications for high energy collider experiments has been studied in \cite{Espinosa:1993bs,Choi:1993cv,Ham:2004cf,Profumo:2007wc,Espinosa:2011ax,Cline:2012hg,Kotwal:2016tex,Chen:2017qcz}. A variant with a complex singlet (two additional degrees of freedom) has been analyzed in \cite{Jiang:2015cwa,Chiang:2017nmu}. The viability of a SFOEWPT arising from scalars charged under SU(3)$_C$ (including, {\it e.g.}, light stops in the MSSM) is severely constrained by the non-observation of these particles at the Large Hadron Collider (LHC) as well as by the measured Higgs boson signal strengths \cite{Katz:2014bha,Katz:2015uja}. 

The constraints on colorless electroweak multiplets are considerably weaker. Here, we consider the colorless electroweak multiplet containing the fewest degrees of freedom, the real triplet $\Sigma$ that has vanishing hypercharge. The collider phenomenology and EWPT dynamics of the \lq\lq $\Sigma$SM" have been considered in Refs.~\cite{FileviezPerez:2008bj,Patel:2012pi}.  The finite-$T$ phase history of the $\Sigma$SM includes the possibility of two-step EWSB, where -- prior to entering the Higgs phase -- the universe enters a phase of broken electroweak symmetry involving a non-vanishing vacuum expectation value (vev) for the neutral component of $\Sigma$ but a vanishing neutral Higgs vev. The transition to the $\Sigma$ phase can be strongly first order, a possibility that is presently less constrained phenomenologically than a single-step SFOEWPT to the Higgs phase. The possibility of baryogenesis during the first step of the two-step scenario has been explored in Ref.~\cite{Inoue:2015pza}. For a general analysis of the two-step EWSB scenario, see Ref.~\cite{Blinov:2015sna}.

The viability of a SFOEWPT (at any step) in the $\Sigma$SM or any other BSM scenario must be validated by non-perturbative computations. The foregoing studies in the xSM, cxSM, $\Sigma$SM and even the two-Higgs doublet model have employed perturbation theory\footnote{However see \cite{Andersen:2017ika} for a recent nonperturbative study of the two-Higgs double following a similar methodology to this paper} \cite{Dorsch:2013wja,Basler:2016obg,Dorsch:2017nza,Basler:2017uxn}. General considerations imply that the perturbative expansion formally breaks down in the vicinity of a  phase transition, as the relevant finite-$T$ expansion parameter becomes large in this region. Indeed, the existence of a crossover transition and the presence of a critical point in the SM have only been observed in non-perturbative computations and not in perturbative studies. Nonetheless, perturbative computations in both the SM and MSSM indicate reasonable qualitative if not quantitative agreement with other features of non-perturbative computations, such as the dependence of thermodynamic properties on the underlying model parameters. 

With an eye toward a more robust assessment of the viability of a SFOEWPT (one- or two-step) in the $\Sigma$SM, we present in this paper a first step toward \lq\lq benchmarking" the existing perturbative analyses. We do so in two parts. First, we derive the dimensionally-reduced,  three-dimensional effective field theories (DR3EFT's) that are most amenable to non-perturbative lattice simulations. Depending on the mass of $\Sigma$, we derive matching relations between the EFT parameters and those of the full theory. Assuming the triplet is  heavy or superheavy (defined in Section \ref{sec:dimred} below) where it is integrated out, we utilize the results of existing non-perturbative computations for the DR3EFT in which the Higgs boson is the only dynamical scalar to analyze the nature of the single-step transition to the Higgs phase.  While this case cannot address the viability of the two-step EWSB scenario since the $\Sigma$ has been integrated out, it does provide one arena in which to compare with the corresponding perturbative calculations. Assessing the dynamics of the two-step scenario will require new lattice computations involving dynamical $\Sigma$ fields. 

In the present case, we find that
\begin{itemize}
\item There exist regions of model parameter space for which a one-step transition to the Higgs vacuum can be first order.  They are shown in Figures \ref{fig:NP1}, \ref{fig:NP2}, and \ref{fig:NP3} below.  However, without further information, we are unable to assess the strength of the phase transition relevant for baryogenesis.
\item For a given value of the physical triplet scalar mass, there is a minimum value of the portal coupling that accommodates a first order transition. Below this critical value, EWSB occurs via a crossover transition.
\item The presence of  a first order transition in this regime is associated with a minimum reduction in the rate for the Higgs boson to decay to two photons. 
\item These features of the EWPT dynamics are not accessible using perturbative computations.
\end{itemize}

In the remainder of the paper, we organize our presentation of this analysis as follows.  In section \ref{sec:model} we formulate and summarize the phenomenology of the $\Sigma$SM.  In section \ref{sec:dimred}, we summarize theoretical aspects of dimensional reduction, and obtain various DR3EFT's for the case the $\Sigma$ is a light degree of freedom.   In Section \ref{sec:heavyandsuperheavycase}, the DR3EFT for the case the $\Sigma$ is heavy or superheavy is derived, and numeral results are presented.  We discuss the implications of our findings in section \ref{sec:discuss}.  A listing of matching relations among the various DR3EFTs are provided in the appendices.

\section{Model and phenomenology}
\label{sec:model}
The $\Sigma$SM is formulated by extending the SM with a scalar isotriplet field $\Sigma^a$ carrying zero hypercharge.  In terms of the SM Higgs isodoublet $H$ and the new isotriplet
\begin{equation}
 H = \begin{pmatrix}\phi^+ \\ \frac{1}{\sqrt{2}}(h + i\phi^0)\end{pmatrix}\enspace\text{and}\enspace
\Sigma^a = \begin{pmatrix}\sigma_1\\ \sigma_2 \\ \sigma_3 \end{pmatrix}\,,
\end{equation} 
the scalar sector Lagrangian, with the metric signature $(+,-,-,-)$, reads \cite{FileviezPerez:2008bj,Patel:2012pi}
\begin{equation}
\mathcal{L} = \textstyle (D_\mu H)^\dag(D^\mu H) + \frac{1}{2}(D_\mu\Sigma)^a(D^\mu\Sigma)^a -V(H,\Sigma) \,,
\end{equation}
where the covariant derivatives in terms of the hypercharge and isospin gauge fields $B_\mu$ and $W_\mu^a$ and coupling constants $g'$ and $g$ are given by
\begin{gather}\begin{aligned}
 D_\mu H &= \textstyle (\partial_\mu + \frac{i}{2} g'  B_\mu + i g \frac{\tau^a}{2} W_\mu^a)H\\
 (D_\mu \Sigma)^a &= (\partial_\mu\delta^{ac} - g  \epsilon^{abc} W_\mu^b)\Sigma^c\,,
\end{aligned}\end{gather}
and the scalar potential is
\begin{multline}
\textstyle V(H,\Sigma) = -\mu^2 H^\dag H+\lambda (H^\dag H)^2-\frac{1}{2}\mu_\Sigma^2(\Sigma^a\Sigma^a)\\
\textstyle +\frac{1}{2}a_2 H^\dag H\Sigma^a\Sigma^a + \frac{1}{4}b_4(\Sigma^a \Sigma^a)^2.
\end{multline}
For simplicity, we have imposed a $Z_2$ symmetry under $\Sigma^a \rightarrow -\Sigma^a$ on the theory that forbids the gauge-invariant cubic portal operator $H^\dag\Sigma^a \frac{\tau^a}{2} H$.   Additionally, we retain only the top quark Yukawa coupling $y_t$ to the SM Higgs doublet, while neglecting all others.

In the potential, we take $\mu^2$ positive so that the neutral Higgs field $h$ obtains a non-zero vacuum expectation value (vev) at sufficiently low temperature ($T$), while for high temperature, thermal corrections change the sign of the quadratic operator, leading to symmetry restoration.  The sign of the triplet quadratic coefficient, $\mu_\Sigma^2$, may be either positive or negative. For $\mu_\Sigma^2 > 0$, the $T=0$ vacuum exhibits several extrema, including minima along the $h$ and $\sigma_3$ directions (for a discussion, see Ref.~\cite{Patel:2012pi}). Here, we focus on the case where the absolute $T=0$ minimum lies along the $h$ direction, with vanishing $\sigma_3$ vev.  In this vacuum, all three components of $\Sigma^a$ are degenerate at leading order, with masses given by
\begin{equation}
\textstyle M_\Sigma^2= -\mu_\Sigma^2+\frac{1}{2} a_2 v^2\,,
\end{equation}
where $v=246$ GeV is the zero-temperature tree-level Higgs vev.  The physical quanta of charged and neutral scalar fields are $\Sigma^\pm = (\sigma_1 \mp i \sigma_2)/\sqrt{2}$ and $\Sigma^0 = \sigma_3$. In what follows, we will express our results in terms of the physical mass $M_\Sigma$ and the portal coupling $a_2$.

The $Z_2$ symmetry $\Sigma\rightarrow-\Sigma$ and the absence of a neutral triplet vev implies that $\Sigma^0$ is stable. For the range of $M_\Sigma$ of interest here ($\sim$ 100 -- 600 GeV), it will contribute a subdominant component of the total dark matter relic density\cite{Cirelli:2005uq}. The corresponding dark matter direct detection constraints on the model parameters can be found in \cite{Chao:2018}.

Additional constraints may arise from searches for new electroweak multiplets at the Large Hadron Collider (LHC).  Due to the $Z_2$ symmetry, electroweak production of $\Sigma^\pm$ and $\Sigma^0$ are expected occur in pairs.  Furthermore, electroweak self energy corrections of $\Sigma$ lead to a small mass splitting between the charged and neutral components by roughly $M_{\Sigma^\pm}-M_{\Sigma^0} \approx 160$ MeV.  Consequently, processes involving the production of $\Sigma^\pm$ will lead to disappearing charge tracks due to its relatively slow decay to $\Sigma^0$ by the emission of a soft pion \cite{FileviezPerez:2008bj}.  Although limits on the existence the charged triplet fermions (\emph{e.g.}, charginos) as a function of the triplet mass and lifetime have been obtained by the ATLAS \cite{Aad:2013yna} and CMS \cite{CMS:2014gxa} collaborations, no significant limits have been placed on the scalar triplet $\Sigma$ due to its much shorter lifetime.  Therefore, the LHC results do not yet  significantly constrain the model parameter space.

\section{Dimensional reduction}
\label{sec:dimred}
In this section we begin the non-perturbative study of the EWPT in the $\Sigma$SM by performing a dimensional reduction to an effective three-dimensional theory.  We start by providing an overview of dimensional regularization, and follow up with its construction as applied to the $\Sigma$SM.  Then we describe the matching procedure and our power counting scheme for relating parameters of the various theories.  Finally, we state our renormalization scheme to numerically determine the values of input parameters.

\subsection{Overview}
\label{sec:overview}
\begin{figure}
\centering
\includegraphics[width=0.5\textwidth]{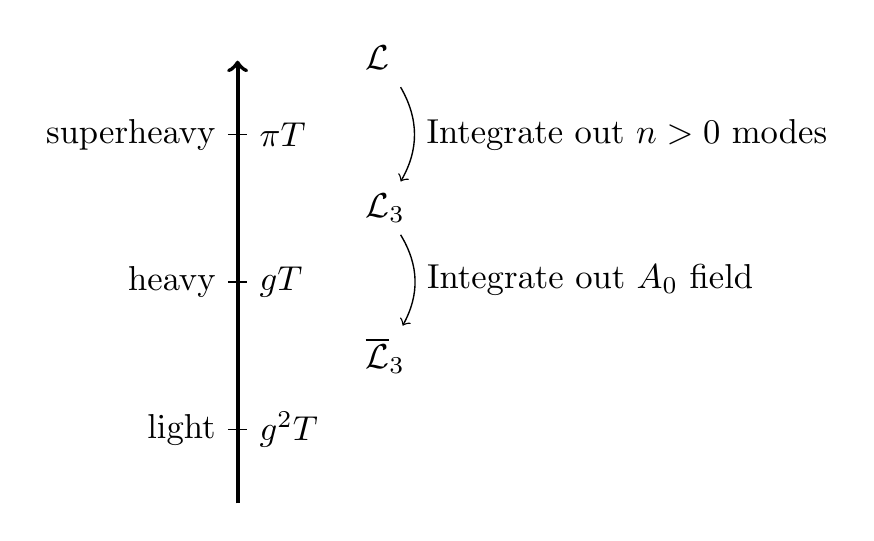}
\caption{Separation of scales in equilibrium thermal field theory in the high-$T$ limit, and the effective theories associated with each scale.  Each effective theory is derived from the one above it by matching Green's functions.}
\label{fig:DR_scales}
\end{figure}
Dimensional reduction is a procedure for constructing an effective three-dimensional theory from the full four-dimensional quantum field theory at a high temperature.  It is made possible by the fact that in the Matsubara formalism of equilibrium thermal field theory, most degrees of freedom decouple from physics in the high temperature limit. 

Following the nomenclature in \cite{Kajantie:1995dw}, the mass scale associated with the lowest non-vanishing Matsubara frequency $\pi T$ is the ``superheavy" scale, so that all Matsubara modes apart from the zero mode of bosonic degrees of freedom are superheavy.  The most prominent dynamical effect of the superheavy modes is to generate thermal masses of order $gT$ for the zero Matsubara modes of scalar fields and time component of the gauge fields.  This dynamically generated scale is called the ``heavy" scale, which is separated from the superheavy scale in the weak coupling limit.  The remaining degrees of freedom --- spatial components of gauge fields --- are ``light" degrees of freedom.

However, scalar fields whose bare mass term is negative, such as the $-\mu^2 H^\dag H$ term of the Higgs isodoublet, will have smaller effective thermal masses due to a cancellation between the bare and the thermally generated ones.  At temperatures around the phase transition where their thermal expectation values are expected to change, the cancellation will be significant to the extent that these scalar fields have effective masses that are far below the heavy scale.  Therefore, the zero Matsubara modes of these scalar fields are also classified as light degrees of freedom.

This hierarchy of scales in the high-$T$ limit is illustrated in Fig. \ref{fig:DR_scales}, and motivates us to pass through a series of three-dimensional effective field theories, ultimately obtaining a DR3EFT involving just the light DOF which is most readily simulated on the lattice for a non-perturbative study of the EWPT.  In the next section we explain how the effective theories are constructed for the $\Sigma$SM.

\subsection{Dimensional reduction to the effective theory at the heavy scale}
We begin our construction of the DR3EFTs by first considering the case $\mu_\Sigma^2 > 0$, so that the zero Matsubara mode of the real triplet $\Sigma$ is classified as a light degree of freedom.  This accommodates the possibility for the real triplet to actively participate in the EWPT with a varying thermal expectation value.  The case where $\mu_\Sigma^2 < 0$, so that it is classified as heavy or superheavy, will be treated in Section \ref{sec:heavyandsuperheavycase} below.

We start by integrating out the non-zero Matsubara modes (superheavy DOF) to obtain a dimensionally reduced effective theory at the heavy scale involving just the zero Matsubara modes.
The most general super-renormalizable euclidean Lagrangian $\mathcal{L}_3$ consistent with the symmetries of the original theory is
\begin{multline}\label{lag_heavy}
\mathcal{L}_3 = \textstyle \frac{1}{4}B_{ij}B_{ij} + \frac{1}{4}W_{ij}^a W_{ij}^a  + \mathcal{L}_\text{3,gf} + (\vec{D}H^\dag)\!\cdot\!(\vec{D} H)  \\
\textstyle + \frac{1}{2} (\vec{D}\Sigma)^a\!\cdot\!(\vec{D}\Sigma)^a 
+ V_3(H,\Sigma) + \mathcal{L}_\text{3,time}\,.
\end{multline}
The first few terms resemble the Lagrangian of the underlying four-dimensional theory.  The hypercharge and isospin field strength tensors are
\begin{gather}\label{covderiv_heavy}
\begin{aligned}
B_{ij} &= \nabla_i B_j - \nabla_j B_i \\
W_{ij}^a &= \nabla_i W_j^a - \nabla_j W_i^a - g_3 \epsilon^{abc} W^b_i W^c_j\,,
\end{aligned}
\end{gather}
the gauge fixing and SU(2) ghost lagrangian is
\begin{equation}
\textstyle \mathcal{L}_{3,\text{gf}} = \frac{1}{2\xi}(\vec\nabla\!\cdot\!\vec{B})^2 + \frac{1}{2\xi}(\vec\nabla\!\cdot\!\vec{W}^a)^2 + (\vec\nabla\eta^a)\!\cdot\!(\vec{D}\eta)^a\,,
\end{equation}
the covariant gradients are
\begin{gather}\begin{aligned}
 \vec{D} H &= \textstyle (\vec\nabla + \frac{i}{2} g'_3  \vec{B} + i g_3 \frac{\tau^a}{2} \vec{W}^a)H\\
 (\vec{D}\Sigma)^a &= (\vec\nabla\delta^{ac} - g_3  \epsilon^{abc} \vec{W}^b)\Sigma^c\,,
\end{aligned}\end{gather}
and the scalar potential is
\begin{multline}\label{lag_heavy_potential}
V_3(H,\Sigma) = \textstyle + \mu^2_3 H^\dag H+\lambda_3 (H^\dag H)^2+\frac{1}{2}\mu_{\Sigma,3}^2(\Sigma^a\Sigma^a) \\
\textstyle + \frac{1}{2}a_{2,3} H^\dag H\Sigma^a\Sigma^a + \frac{1}{4}b_{4,3}(\Sigma^a \Sigma^a)^2\,.
\end{multline}
Additionally, due to the absence of full Lorentz invariance of the theory at finite temperature, additional terms arise in the effective theory involving the time component of gauge fields,
\begin{multline}
\label{lag_heavy_time}
\textstyle
\mathcal{L}_\text{3,time}={\textstyle\frac12[(\vec\nabla W_0^a)^2+m_D^2(W^a_0)^2] +\frac14\kappa_3(W^a_0W^a_0)^2} \\
{\textstyle+\frac12[(\vec\nabla B_0)^2 +m_D'^2B_0^2] +\frac14\kappa_3'B_0^4  +\frac14\kappa_3''(W^a_0)^2 B_0^2}\\
+ \textstyle\frac12[(\vec\nabla G^A_0)^2 + m''^2_D (G^A_0)^2] \\
+h_{3}H^\dag H  (W^a_0)^2 +h_{3}'H^\dag H  B_0^2 +h_{3}''B_0 H^\dagger (W_0^a\tau^a)H \\ 
+ \omega_{3} H^\dag H  (G^A_0)^2  + \delta_3 (\Sigma^a)^2(W^b_0)^2 + \delta_3' (\Sigma^a W^a_0)^2.
\end{multline}
Since the effect of gluons fields $G^A_0$ and $\vec{G}^A$ and the associated SU(3) ghosts arise  through top quark loops, only those interaction terms involving them that are needed in subsequent calculations are explicitly displayed above for brevity (see section 2.2.3 of Ref.~\cite{Brauner:2016fla}).

Formulae connecting the coupling constants and normalization of the fields in $\mathcal{L}_3$ to the couplings and zero Matsubara modes of the full four dimensional theory in $\mathcal{L}$ are obtained by matching, to be discussed in more detail below in Section \ref{sec:matching_example}, and are listed in Appendix \ref{sec:matching_1loop}.

\subsection{Reduction to the theory at the light scale}
As explained above, the effect of integrating out the nonzero Matsubara modes at the superheavy scale is to induce thermal masses of scalar fields and time component of gauge fields of order $gT$, which in the weak coupling limit are separated from the superheavy scale but in the high temperature limit are separated from the light scale.  Continuing with our assumption that $\Sigma$ is light, the only degrees of freedom at the heavy scale that need to be integrated out to obtain an effective theory at the light scale are the time component of gauge fields $B_0$, $W_0^a$ and $G_0^A$.

The most general super-renormalizable effective Lagrangian involving the light degrees of freedom is
\begin{multline}\label{lag_light}
\bar{\mathcal{L}}_3 = \textstyle \frac{1}{4}B_{ij}B_{ij} + \frac{1}{4}W_{ij}^a W_{ij}^a + \bar{\mathcal{L}}_{3,\text{gf}} + (\vec{D}H^\dag)\!\cdot\!(\vec{D} H)\\
\textstyle + \frac{1}{2} (\vec{D}\Sigma)^a\!\cdot\!(\vec{D}\Sigma)^a 
+ \bar{V}_3(H,\Sigma)\,,
\end{multline}
with the same abbreviations listed in (\ref{covderiv_heavy})--(\ref{lag_heavy_potential}), but with new couplings which we distinguish with a bar: $\bar{g}_3$, $\bar{g}'_3$, $\bar{\lambda}_3$, $\bar{a}_{2,3}$, etc.
The form of the effective Lagrangian at the light scale is identical to that at the heavy scale in (\ref{lag_heavy}), but without $\mathcal{L}_{3,\text{time}}$.  The relations connecting the coupling constants at the heavy scale and the light scale are derived by matching are listed in Appendix \ref{sec:matching_light}.  

\subsection{Matching of the parameters}
\label{sec:matching_example}

In this subsection, we explain how the field normalizations and coupling constants between sets of EFTs are derived.  Additionally, we specify our power counting scheme and the level of precision we derive these matching relations.  For details of the matching procedure, see also Refs.~\cite{Kajantie:1995dw} and \cite{Brauner:2016fla}.

We adopt a power counting scheme similar to that of Ref.~\cite{Kajantie:1995dw} wherein the quartic couplings $\lambda$, $a_2$, and $b_4$ scale as the square of the SU(2) gauge coupling constant $g^2$, while the top quark Yukawa $y_t$ and the remaining gauge coupling constants $g'$,$g_s$ scale linearly with $g$.  Additionally, under the assumption that both the Higgs doublet $H$ and the real triplet $\Sigma$ are light, the negative mass parameters $\mu^2$ and $\mu_\Sigma^2$ are required to scale as $g^2 T^2$ near the electroweak phase transition, as explained in Section \ref{sec:overview} above.  We perform dimensional reduction perturbatively, in the symmetric phase in the Fermi-$\xi$ gauges, to order $O(g^4)$.  This requires the evaluation of self energy functions through two loop to match mass parameters $\mu^2$ and $\mu_\Sigma^2$, one loop diagrams to match the remaining coupling constants.  To illustrate how the matching relations for fields and couplings between the EFTs are derived, we summarize the procedure, using as the portal couplings $a_{2,3}$ and $\bar{a}_{2,3}$ as a representative example.

The formula for $a_{2,3}$ listed in (\ref{portal_heavy}), is determined by requiring that the four point Green's function $\langle H^{\dagger}H \Sigma^a\Sigma^a\rangle$ in both the 4d theory and the heavy scale 3d theory match at the matching scale $\Lambda$.  This is possible provided the fields in the 3d theory are canonically normalized.  Canonical normalization is achieved by comparing the two-point Green's function in the two theories.  For a generic bosonic field $\phi$, this relationship reads
\begin{equation}
\phi_\td^2 = \frac{1}{T}\big[1 + {\hat\Pi}_{\phi}'(0, 0)\big] \phi^2,
\label{fieldmatching}
\end{equation}
where ${\hat\Pi}_{\phi}(\omega^2, \vec{p}^2)$ is the fully renormalized self-energy function of the Matsubara frequency $\omega$ and spatial momentum $\vec{p}$, and the prime denotes a derivative with respect to $\vec{p}^2$.  The explicit factor of $1/T$ accounts for absorbing a similar factor in front of the 3D effective Lagrangian.  To ultimately obtain an $\mathcal{O}(g^4)$ accuracy in the matching relations, (\ref{fieldmatching}) needs to be known only to one loop order.  Additionally, only contributions from the $n \neq 0$ Matsubara modes should be included.

The portal coupling $a_{2,3}$ can be then be determined by comparing the corresponding tree-level  vertex in the DR3EFT against the one in the 4d calculated to $O(g^4)$.  The 3d vertex reads
\begin{equation} \label{tree_portal_heavy}
-a_{2,3} T (H^{\dagger}H\Sigma^a\Sigma^a)_{\text{3d}},
\end{equation}
where the $T$ follows from the rescaling of the 3d fields.  The corresponding vertex in the 4d theory is 
\begin{equation} \label{loop_portal_superheavy}
(-a_2 - \hat\Gamma(0)) (H^{\dagger}H\Sigma^a\Sigma^a)_{\text{4d}}
\end{equation}
where ${\hat\Gamma}(0)$ is the connected (fully-renormalized) one-loop $H^{\dagger}H\Sigma^a\Sigma^a$ vertex function at zero external momentum and excluding the zero Matsubara modes.  By matching (\ref{tree_portal_heavy}) and (\ref{loop_portal_superheavy}), and accounting for the difference in the field normalization in (\ref{fieldmatching}), we obtain the desired matching formula for the portal coupling
\begin{equation}
a_{2,3} = T \big[a_2 - a_2({\hat\Pi}_{H}'(0)+{\hat\Pi}_{\Sigma}'(0))+\hat\Gamma(0)\big]\,.
\end{equation}
All other matching relations between the superheavy and heavy scales listed in Appendix \ref{sec:matching_1loop} are derived in a similar way, using the table of integrals found in \cite{Kajantie:1995dw}.  To minimize logarithms, and to eliminate factors $\ln(4\pi)-\gamma_E$, we choose the matching scale to be $\Lambda = 4\pi T/e^{\gamma_E}$.

To obtain the portal coupling $\bar{a}_{2,3}$ at the light scale, where the time component of gauge fields $B_0$, $W_0^a$, $G_0^A$ are integrated out, an analogous procedure is followed.  Field and mass parameters are again related by comparing self energy functions.  However, there is no change in normalization of the scalar fields in the two theories as there are no contributions giving momentum dependence.  This leads to the simpler matching relation
\begin{equation}
\bar{a}_{2,3} = a_{2,3} + \hat\Gamma_{3}(0),
\end{equation}
where $\hat\Gamma_{3}(0)$ is the contribution from the $B_0$, $W_0^a$ and $G^A_0$ fields to the $H^{\dagger}H\Sigma^a\Sigma^a$ connected Green's function in the ``high" scale DR3EFT.

It is worth highlighting one technical point appearing at two-loop order matching of the mass parameters. Since the effective 3d theories are super-renormalizable due to the reduced number of spacetime dimensions, running of the 3d parameters can be solved exactly at the two-loop level. In particular, the couplings are manifestly independent of the RG scale, and renormalization is only needed for mass parameters at two-loop \cite{Farakos:1994kx}. On dimensional grounds, the renormalized mass parameters are of the form 
\begin{align}
\label{eq:3d-running}
\mu_3^2 = f_3 \ln \frac{\Lambda_0}{\Lambda_{\text{3d}}},
\end{align}
where $f_3$ is an $O(g_3^4)$ function of the 3d couplings, $\Lambda_{3d}$ is the RG scale of the 3d theory and $\Lambda_0$ is a mass scale that is determined by the matching procedure. Note that $f_3$ corresponds to the mass counterterm $\delta\mu_3^2$, since the bare mass, defined as $\mu_{3(b)}^2 = \mu_3^2 + \delta\mu_3^2$, has to be RG invariant.

The 3d bare mass is also independent of the renormalization scale of the 4d theory. In the $O(g^4)$ matching relation for $\mu^2_3$, there is a logarithmic term of the form 
\begin{equation}
T^2 f_4 \ln \frac{\Lambda_0}{\Lambda_{\text{4d}}},
\end{equation}
where $f_4$ is a function of the 4d couplings that matches $f_3$ to $O(g^4)$ accuracy, and $\Lambda_{4d}$ is the 4d RG scale. This term cancels the $\Lambda_{\text{4d}}$ dependence coming from the 3d mass counterterm, expressed in terms of 4d parameters to order $O(g^4)$. In particular, the constant $\Lambda_0$ can be calculated in the 4d theory and equals $3T e^c$, where 
\begin{equation}
c \equiv \frac{1}{2} \bigg( \ln\big( \frac{8\pi}{9}\big) + \frac{\zeta'(2)}{\zeta(2)} - 2 \gamma_E\bigg)
\end{equation}
is a constant appearing naturally from two-loop thermal sum-integrals.

We may now replace this logarithmic term in the mass parameter matching relation by the more accurate running of the 3d mass in Eq.~\ref{eq:3d-running}, which receives no corrections at higher loop orders due to the super-renormalizable nature of the effective theory. This is the reason for the appearance of 3d parameters in the two-loop matching relations in Appendix~\ref{sec:matching_1loop}.

\subsection{Renormalization and the numerical determination of parameters}
\label{sec:renormalization}
For a numerical study of the phase diagram in this model, it remains to fix the input parameters of the underlying model at the superheavy scale.  The theory depends on 5 parameters of the SM, $\mu^2$, $\lambda$, $g'$, $g$, and $y_t$, together with 3 additional parameters from the extended sector, $\mu_\Sigma^2$, $a_2$ and $b_4$.  We determine their values in the $\overline{\text{MS}}$ scheme by relating them to measured observables.

We choose to fix $\mu^2$, $\lambda$, $g'$, and $g$, by relating them to the fine structure constant $\hat\alpha(M_Z^2)$ and the poles masses $M_W$, $M_Z$, $M_H$, at the scale $\Lambda=M_Z$.  Although $G_F$ is conventionally used in place of $M_W$ for a more precise determination, at the level of precision we are working, we choose to work with use $M_W$ for clarity.  In terms of the Higgs self energy function $\Sigma_H$ and the transverse polarization functions of the gauge bosons $\Pi_W$ and $\Pi_Z$, the one loop relations are
\begin{align}
\mu_H^2 &= \frac{M_H^2}{2}\Big(1-\frac{\Sigma_H (M_H^2)}{M_H^2}\Big)\\
\nonumber\lambda &= \frac{\pi \hat\alpha M_H^2 M_Z^2}{2M_W^2 (M_Z^2 - M_W^2)}\Big[1 - \frac{\Sigma_H (M_H^2)}{M_H^2} - \frac{\Pi_Z(M_Z^2)}{M_Z^2}\\
& \qquad +\frac{\Pi_W(M_W^2)}{M_W^2} + \frac{\Pi_Z(M_Z^2) - \Pi_W(M_W^2)}{M_Z^2 - M_W^2}\Big]\\
g'^2 &= \frac{4\pi\hat\alpha M_Z^2}{M_W^2}\Big[1 - \frac{\Pi_Z(M_Z^2)}{M_Z^2} + \frac{\Pi_W(M_W^2)}{M_W^2}\Big] \\
\nonumber g^2 &= \frac{4\pi\hat\alpha M_Z^2}{M_Z^2 - M_W^2}\Big[1 - \frac{\Pi_Z(M_Z^2)}{M_Z^2} \\
& \hspace{3cm} + \frac{\Pi_Z(M_Z^2)-\Pi_W(M_W^2)}{M_Z^2-M_W^2}\Big]\,.
\end{align}
The relationship for the top quark Yukawa coupling additionally depends on its self energy function, parametrized in terms of invariant functions as 
\begin{equation*}
-i\Sigma(\slashed{p}) = -i(\slashed{p}\,A(p^2) + M_t\,B(p^2)).
\end{equation*}
At one loop order, the relationship is
\begin{align}
\nonumber y_t^2 &= 2\pi \hat\alpha \frac{M_Z^2 M_t^2}{M_W^2(M_Z^2 - M_W^2)}\Big[1 - \frac{\Pi_Z(M_Z^2)}{M_Z^2} + \frac{\Pi_W(M_W^2)}{M_W^2} \\
 & + \frac{\Pi_Z(M_Z^2)-\Pi_W(M_W^2)}{M_Z^2-M_W^2} - 2(A(M_t^2) + B(M_t^2))\Big]\,.
\end{align}
which we use to fix the Yukawa coupling at the scale $\Lambda=M_t$.

Finally, among the three parameters of the extended sector, we only choose to express the mass parameter of the real triplet $\mu_\Sigma^2$ in terms of the physical pole mass of the electrically neutral triplet $\Sigma^0$ at the scale $\Lambda=M_\Sigma$.  In terms of the neutral triplet self energy function $\Sigma_\Sigma$, the one loop relationship is given by
\begin{multline}
\mu_\Sigma^2 = - M_\Sigma^2 + \Sigma_\Sigma(M_\Sigma^2) + \frac{a_2 M_W^2}{2 \pi \hat\alpha}\Big(1-\frac{M_W^2}{M_Z^2}\Big)\times \\ 
\Big[1-\frac{\Pi_W(M_W^2)}{M_W^2} - \frac{\Pi_W(M_W^2)/M_W^2 - \Pi_Z(M_Z^2)/M_Z^2}{1-M_Z^2/M_W^2}\Big]\,.
\end{multline}
Since no meaningful measurements have been made to fix the remaining parameters $a_2$ and $b_4$, in what follows, we will present our results in terms of them directly at the scale $\Lambda=M_Z$.

Having determined the values of renormalized parameters at their chosen scales, we solve the one loop renormalization group equations to obtain their values at the matching scale $\Lambda = 4\pi T /e^\gamma_E$
\begin{align}
\label{running_g}
\Lambda \frac{d\,g^2}{d\Lambda} &= - \frac{g^4}{8 \pi^2} \bigg( \frac{22}{3}-\frac{N_d+2 N_t}{6}-\frac{4}{3}N_f \bigg), \\
\Lambda \frac{dg'^2}{d\Lambda} &= \frac{g'^4}{8 \pi^2} \bigg( \frac{N_d}{6} + \frac{20}{9}N_f \bigg), \\
\Lambda \frac{d\,y_t^2}{d\Lambda} &= \frac{y_t^2}{8 \pi^2} \bigg( \frac{9}{2}y_t^2  - \frac{9}{4}g^2 - \frac{17}{12}g'^2 - 8 g^2_s \bigg), \\
\nonumber \Lambda \frac{d\,\mu^2}{d\Lambda} &= \frac{1}{16\pi^2} \bigg(-3\mu^2\Big(\frac{3}{2}g^2 +\frac{1}{2}{g'}^2 - 2 y_t^2 - 4 \lambda\Big) \\
& \qquad +  3\mu^2_\Sigma a_2  \bigg), \\
\Lambda \frac{d\,\mu^2_{\Sigma}}{d\Lambda} &= \frac{1}{16\pi^2} 2\bigg(2a_2\mu^2 - 6g^2\mu^2_{\Sigma} + 5 b_4 \mu^2_\Sigma \bigg), \\
\nonumber \Lambda \frac{d\,\lambda}{d\Lambda} &= \frac{1}{16\pi^2} \frac{1}{2} \bigg(48\lambda^2_1 +3a^2_2 +\frac{3}{4}(3g^4 + {g'}^4 + 2 g^2{g'}^2)\\
&  - 12 y_t^4 -6\lambda (3g^2 + {g'}^2 - 4 y_t^2) \bigg), \\
\nonumber \Lambda \frac{d \, a_2}{d\Lambda}&= \frac{1}{16\pi^2} 2\bigg(a_2 \Big(-\frac{33}{4}g^2 - \frac{3}{4}{g'}^2 + 3y_t^2 \\
& \qquad + 2 a_2 + 5 b_4 + 6 \lambda \Big) + 3 g^4 \bigg), \\
\Lambda \frac{d\,b_4}{d\Lambda} &= \frac{1}{16\pi^2} 2\bigg( -12b_4 g^2 + a^2_2 + 11 b^2_4 + 6g^4 \bigg)\,.
\end{align}
By allowing the parameters in the tree level Lagrangian to vary with the renormalization scale, we have observed that our results exhibited reduced sensitivity to the chosen value of $\Lambda$.

Having derived the DR3EFT at the light scale (\ref{lag_light}) and established a renormalization scheme to fix the input parameters, the next step is to perform a non-perturbative numerical study of this theory on the lattice.  We postpone the lattice formulation of this theory, together with a comparison of numerical results with perturbation theory to Part II of this series.  Instead, in the next section, we turn to the case where the real triplet $\Sigma$ is either heavy or superheavy, for which we can use existing lattice results to study the EWPT non-perturbatively.

\section{Heavy and Superheavy Triplet}
\label{sec:heavyandsuperheavycase}
In the case $\mu_\Sigma^2<0$, the real triplet degrees of freedom $\Sigma$ are either at the heavy over superheavy scales.   This implies that it is integrated out in first or second step of dimensional reduction, and is absent from the DR3EFT at the light scale.  Although this assumption precludes the possibility of $\Sigma$ changing its thermal expectation value of during the EWPT, the resulting DR3EFT is of the same form as that obtained from the minimal SM,
\begin{equation}
\bar{V}_3(H) = \bar{\mu}^2_{3} H^{\dagger} H + \bar{\lambda}_3(H^{\dagger}H)^2\,,
\end{equation}
but where the influence of the heavy or superheavy $\Sigma$ is encoded in the matching relations listed in Appendices~\ref{sec:heavy_triplet} and \ref{sec:matching_superheavy}, respectively.
Since the thermodynamics of the EWPT of this theory has previously been studied on the lattice, numerical simulations of this theory have already been \cite{Kajantie:1995dw}, we may readily apply the results in this case to study the EWPT in the $\Sigma$SM.

Properties of the EWPT on the lattice are characterized by two temperature-dependent dimensionless parameters
\begin{equation}
x = \frac{\bar{\lambda}_3}{\bar{g}_3^2}, \quad y = \frac{\bar{\mu}_{3}^2}{\bar{g}_3^4}.
\end{equation}
The results of the simulations are as follows.  The critical temperature occurs near where the $y$ parameter changes sign; when $x$ is sufficiently small but positive \mbox{$0 < x \lesssim 0.11$}, the EWPT is first order \cite{Rummukainen:1998as}.  At $x\approx0.11$ the system exhibits a second order EWPT, and for larger values of $x$, the transition is a crossover.  We note that the upper bound on $x$ has been obtained using 3d lattice results for the SU(2) plus Higgs theory and allowing for a $\sim 10\%$ correction from neglected U(1)$_Y$ contributions.  In Section \ref{sec:one_step_results}, we will present our results based on the numerical analysis for the case that $\Sigma$ is a heavy degree of freedom.  We make a comparison with the superheavy case in Appendix \ref{sec:matching_superheavy}.

\subsection{On the validity of dimensional reduction}
Following \cite{Kajantie:1995dw}, we can check the validity of the dimensional reduction by estimating the impact of the higher-dimensional operators that have been dropped from the  light scale DR3EFT on the vevs of the scalars in the effective theory. 

The lowest dimension operators omitted from the heavy and light scales are the (marginal) dimension-three operators $\Lambda_6(H^\dag H)^3_\td$ and $\bar{\Lambda}_6(H^\dag H)^3_\td$, respectively.  Upon integrating out the superheavy scale, the coefficient of the operator at the heavy scale is
\begin{multline}
\Lambda_{6} = \frac{\zeta(3)}{16384\pi^4} \bigg( 3g^6 + {g'}^6 \nonumber \\
+ 3g^2 {g'}^2(g^2 + {g'}^2) +640\lambda^3 - 224 y_t^6 + 8 a_2^3 \bigg)\,.
\end{multline}
The top quark contribution dominates over other SM contributions. The dominant correction $\bar{\Lambda}_6$ in the $\Sigma$SM comes not from the superheavy scale, but from the second step of DR when the heavy triplet is integrated out. The total dimension-three coefficient can be written as 
\begin{align}
\bar{\Lambda}_6 = \Lambda_6 + \Lambda_6^{\text{heavy}},
\end{align}
and the $\Sigma$SM contribution to $\Lambda_6^{\text{heavy}}$ is 
\begin{align}
\Lambda^{\text{heavy}}_{6} (\Sigma) = \frac{1}{512\pi} \left(\frac{a_{2,3}}{\mu_{\Sigma,3}}\right)^3.
\end{align} 
Note that the time component of gauge fields have a subdominant effect when integrating out the heavy scale \cite{Kajantie:1995dw}. 

The top quark contribution 
\begin{align}
\Lambda_{6}(t) &=-\frac{7\zeta(3)}{512\pi^4}y^6_t
\end{align}
shifts the position of the Higgs vev by about one percent in the pure SM. We can estimate the effect of the dimension-three $(H^\dag H)^3$-operator by comparing the magnitude of the dominant $\Sigma$SM contribution to that of the top quark. If the ratio 
\begin{align}\label{delta6}
\Delta_6 \equiv \left| \frac{\Lambda^{\text{heavy}}_{6} (\Sigma) }{\Lambda_{6}(t)} \right|
\end{align}
becomes large, the accuracy, and eventually the validity, is DR3EFT compromised.

\subsection{Results for one-step transition with superheavy and heavy triplet}
\label{sec:one_step_results}

\begin{figure}
\begin{center}
\includegraphics[width=0.5\textwidth,clip=true]{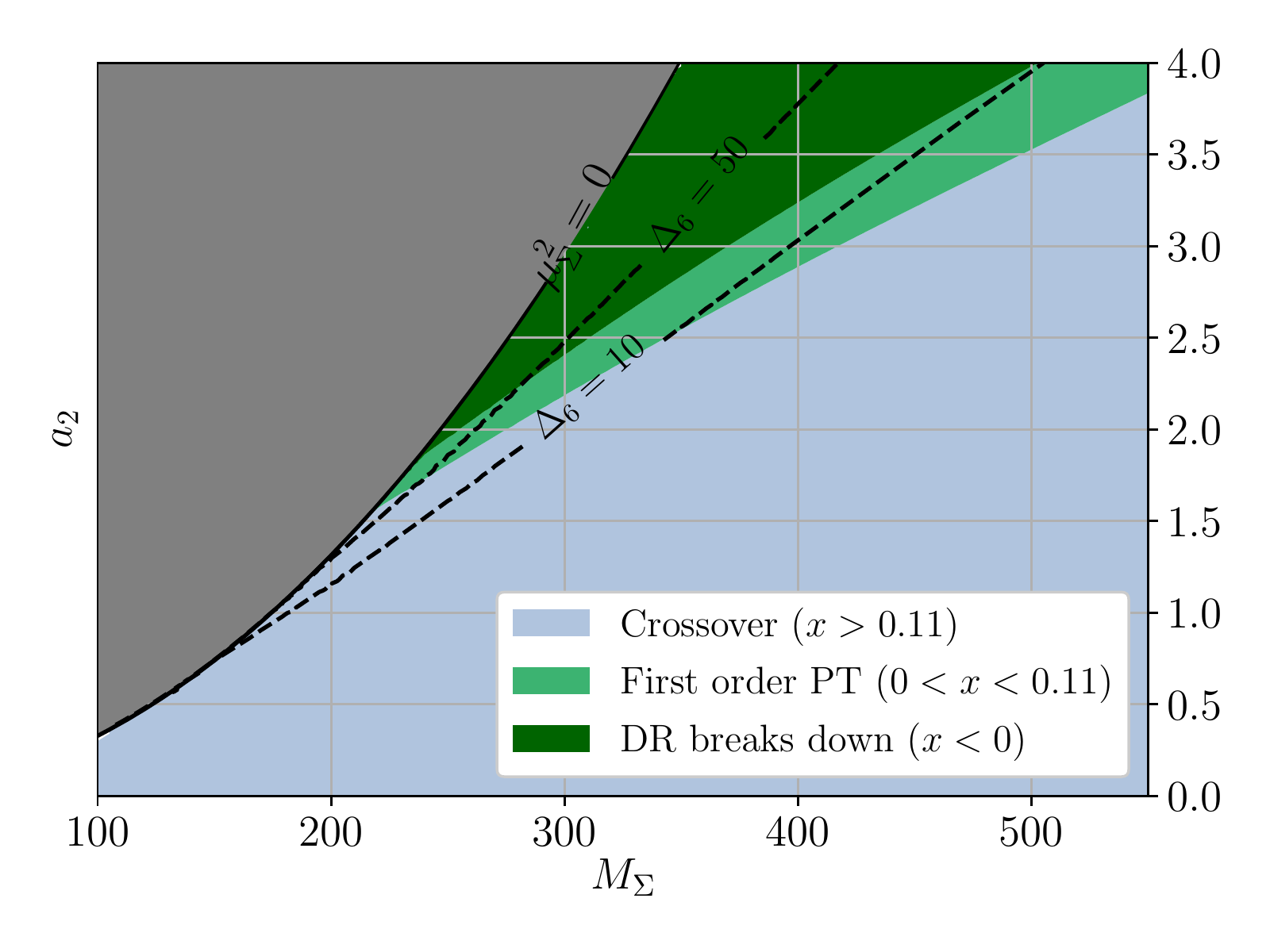}
\end{center}
\caption{$\Sigma$SM phase diagram for a heavy triplet as a function of the triplet mass $M_\Sigma$ and the portal coupling $a_2$. Light blue and light green regions correspond to a one step cross over and first order EWSB transition to the Higgs vacuum, respectively, starting from the electroweak symmetric phase at high $T$. In the the dark green region, assumptions of dimensional reduction (DR) no longer hold. The gray region, corresponds to $\mu_\Sigma^2>0$ where the real triplet is expected to participate in EWPT and must be classified as a light degree of freedom.  Therefore this area of parameter space needs inclusion of its dynamics in the Monte Carlo simulations, so no statement about the phase structure is made here. The dotted lines indicate contours of constant $\Delta_6$ defined in (\ref{delta6}).}
\label{fig:NP1}
\end{figure}

\begin{figure}
\begin{center}
\includegraphics[width=0.5\textwidth,clip=true]{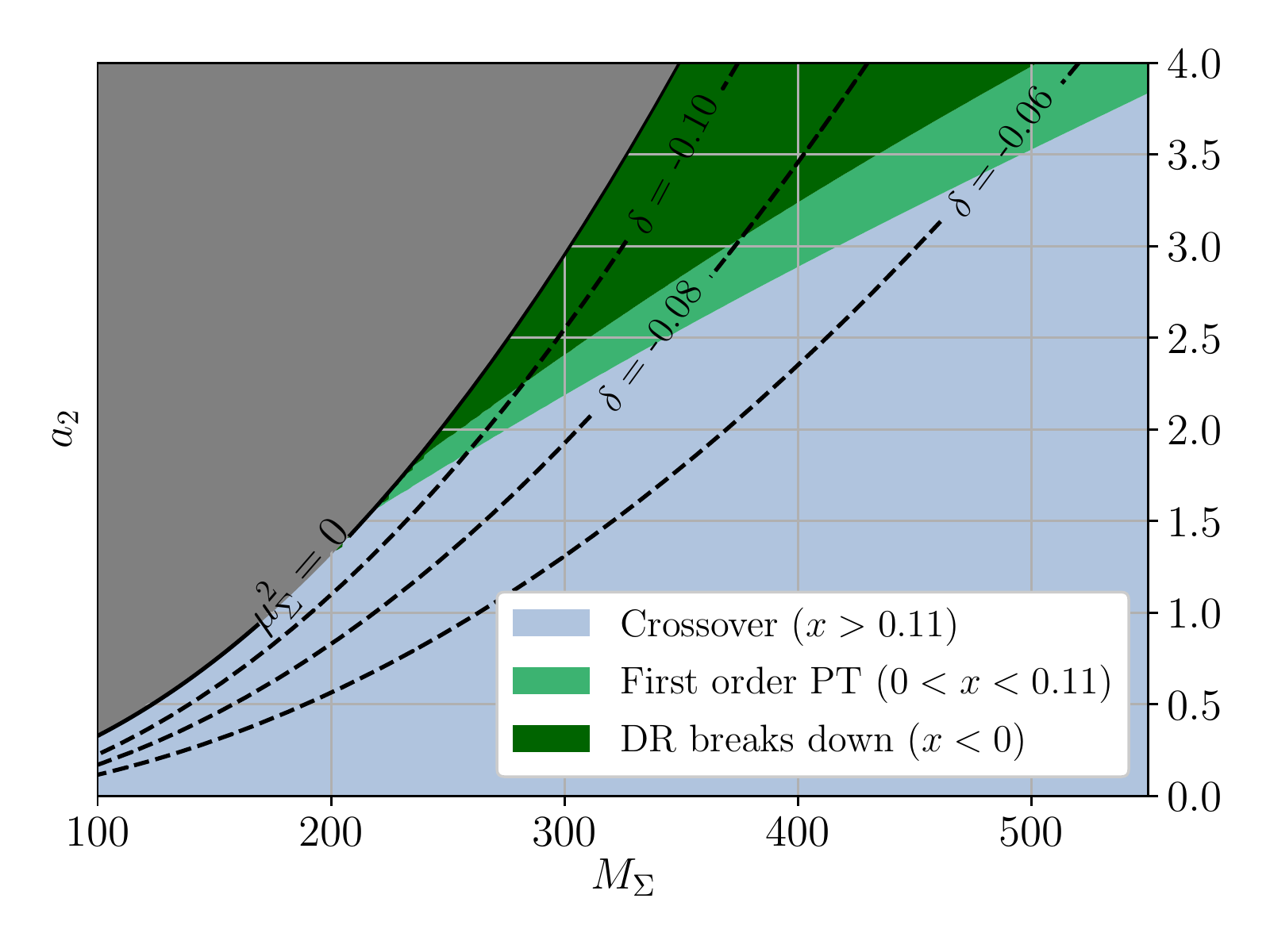}
\end{center}
\caption{Same as Fig.~\ref{fig:NP1} but showing the relative change $\delta$ in the partial width $\Gamma(h\to\gamma\gamma)$, defined in (\ref{eq:diphotonchange}). Dashed lines indicate contours of constant $\delta$ for regions of the parameter space relevant to this analysis.}
\label{fig:NP2}
\end{figure}

\begin{figure}
\begin{center}
\includegraphics[width=0.5\textwidth,clip=true]{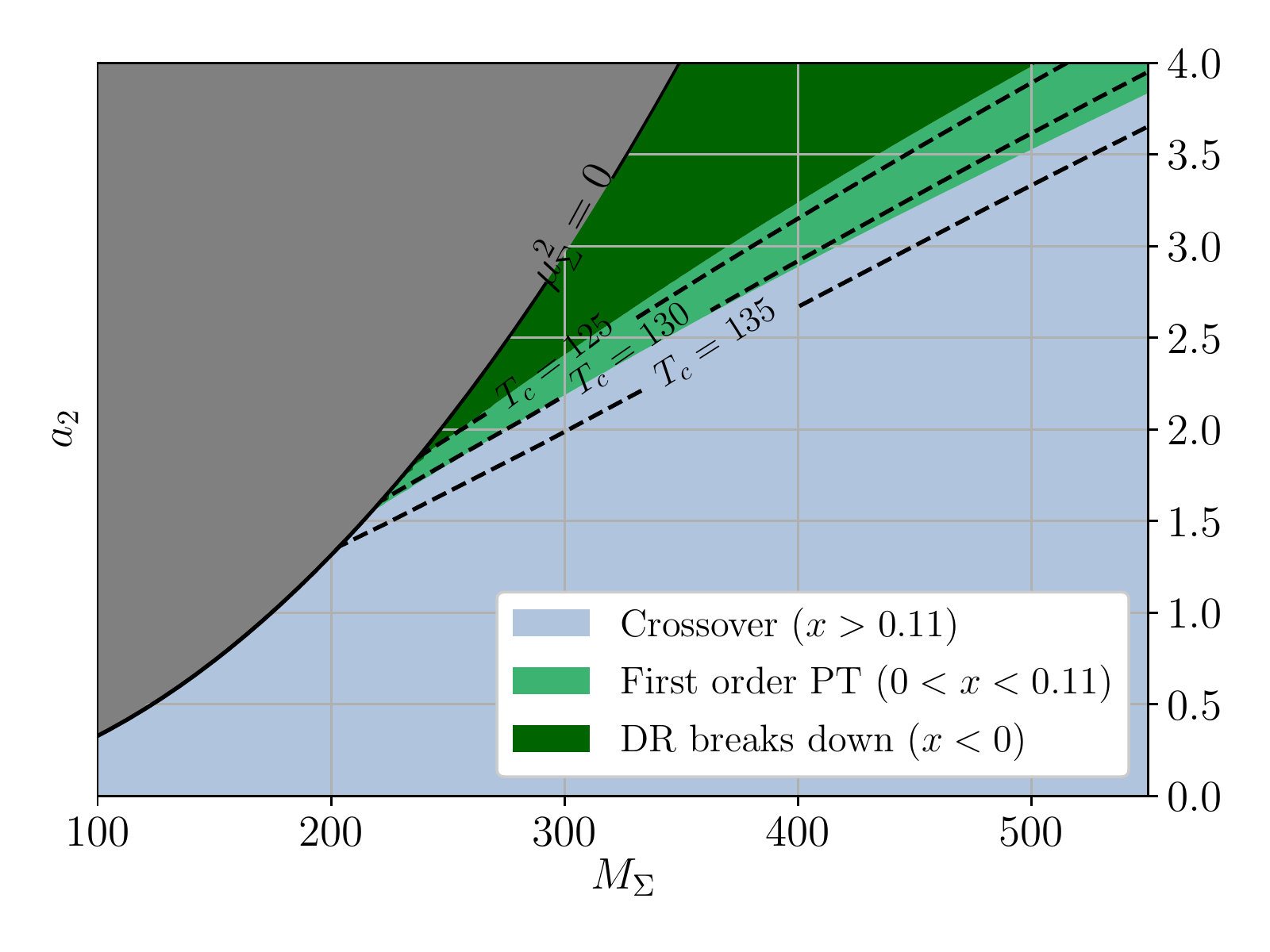}
\end{center}
\caption{Same is Fig.~\ref{fig:NP1} but showing values of the critical temperature. Dashed lines show contours of constant $T_c$ in the vicinity of the first order one-step transition.}
\label{fig:NP3}
\end{figure}

With the foregoing DR3EFT set-up for the heavy $\Sigma$ and matching conditions in hand, we map out the phase diagram for the theory in the ($M_\Sigma$, $a_2$) plane by scanning over the parameters of the potential, determining the values of $x$ and $y$, and identifying the region for a first order EWPT as obtained in the study of Ref.~\cite{Rummukainen:1998as}. We have performed this scan over the $(M_\Sigma,a_2)$-parameter space assuming a uniform distribution of the parameters. The triplet mass was varied from 100 to 600 GeV at intervals of 5 GeV, and portal coupling $a_2$ from 0 to 4 at intervals of 0.05. We then omit all the points in which the triplet mass parameter squared is positive according to tree-level relation.

For each point, we scan the temperature from 80 to 200 GeV at intervals of 20 GeV, and find the critical temperature $T_c$ by interpolation from the condition that $y=0$. To obtain the phase diagram, we determine the value of $x(T_c)$ at  each point in the parameter space. For purposes of visualization, we perform a linear interpolation to obtain contours of constant $x$. Note that a resolution of the uniform scan is chosen to be so dense that outcome of the plot does not visibly change if resolution is made finer. We have verified that the values of $x$ and $y$ are not strongly sensitive to choice of triplet self-coupling. In all results discussed below we have fixed $b_4 = 0.75$. 

The results are displayed in Figs. \ref{fig:NP1}, \ref{fig:NP2}, and \ref{fig:NP3}. In each case, we have identified regions where the EWSB transition from the high-$T$ symmetric phase is a one-step crossover or first order transition, corresponding to the light blue and light green regions, respectively. The dark green regions correspond to choices of the parameters for which the validity of the DR3EFT breaks down. The gray regions, above and to the left of the line $\mu_\Sigma^2=0$, indicate regions of parameter space for which one requires inclusion of an explicit $\Sigma$ in the Monte Carlo simulations. Consequently, we make no statement about the phase structure for this region. We anticipate, however, that the two-step transition analyzed perturbatively in Ref.~\cite{Patel:2012pi}, will emerge in this region from the future lattice study of the gray region.

A key feature of each plot is the existence of a choice of parameters giving a first order or a cross over transition as well as the phase boundary between the two situations. We emphasize that one cannot identify the existence of the cross over region and the boundary with the first order region from a purely perturbative analysis. The results given here, thus, underscore the importance of carrying out a non-perturbative study in order to obtain a physically complete and quantitatively realistic picture of the phase structure of the theory. 

Going beyond this primary point, each of Figs. \ref{fig:NP1}, \ref{fig:NP2}, and \ref{fig:NP3} contain a set of dashed curves that highlight various theoretical and phenomenological considerations. The dashed curves in Fig.~\ref{fig:NP1} give contours of constant $\Delta_6$, defined in (\ref{delta6}). Recall that $\Delta_6$ characterizes the relative magnitudes of $\Sigma$ and top-quark contributions to the coefficient of the higher dimensional $(H^\dag H)^3$ operator in the potential. A rough indication of the importance of this operator on the quantities relevant to the phase transition was obtained in Ref.~\cite{Kajantie:1995dw}, where it was shown that the presence of $\Lambda_6(t)$ leads to a one percent shift in the value of the Higgs vev. We would, thus, expect the relative impact of $\Lambda_6^\text{heavy}(\Sigma)$ to scale linearly with the ratio $\Delta_6$. For sufficiently large $a_2$ and light $M_\Sigma$, one would, thus, expect corrections of greater than $\sim 10\%$ in the value of the Higgs vev associated with the triplet-induced higher dimension operators. The value of the Higgs vev itself is, of course, not directly relevant to the boundaries of the phase diagram, the critical temperature, {\it etc.}, but it does provide one way to assess the quantitative impact of theoretical uncertainties. We defer a more complete determination of the corrections from higher dimensional operators on the phase transition properties to future work, and take the contours of constant $\Delta_6$ as rough indications of the accuracy of our present DR3EFT treatment. 

In Fig.~\ref{fig:NP2}, we illustrate the implications of this study for measurements of Higgs boson couplings. Of particular interest is the rate for the decay to two photons, $\Gamma(h\to\gamma\gamma)$. As discussed in detail in Refs.~\cite{FileviezPerez:2008bj,Patel:2012pi}, loops involving the charged components of the triplet will contribute to the di-photon decay rate, shifting its value from the SM prediction as a function of ($M_\Sigma$, $a_2$). Defining the relative shift
\begin{equation}\label{eq:diphotonchange}
\delta = \frac{\Gamma^{\Sigma\mathrm{SM}}(h\to\gamma\gamma)-\Gamma^{\mathrm{SM}}(h\to\gamma\gamma)}{\Gamma^{\mathrm{SM}}(h\to\gamma\gamma)}
\end{equation}
we plot in Fig.~\ref{fig:NP2} contours of constant $\delta$ in the vicinity of the first order transition region and the boundary with the crossover region. Note that in call cases, $\delta<0$. We emphasize that each point along the boundary between the first order and crossover regions corresponds to a minimum value of $|\delta|$. This feature would allow one to exploit a measurement of $\Gamma(h\to\gamma\gamma)$ (or the corresponding branching ratio) to probe the nature of the transition. For fixed $M_\Sigma$, for example, a sufficiently large and negative deviation of the di-photon rate would indicate the existence of a first order transition, whereas a smaller magnitude or positive value for $\delta$ would imply a crossover transition. 

A separate experimental study would be required to identify $M_\Sigma$. Under the assumptions of the study here, wherein $\Sigma^0$ obtains no vev, such a study could include the search for disappearing charge tracks, as discussed in Ref.~\cite{FileviezPerez:2008bj}. One expects the high luminosity phase of the Large Hadron Collider to enable a determination of the di-photon rate with $\sim 5-10\%$ precision\cite{Brandstetter:2018eju}, potentially allowing one to probe the lower $M_\Sigma$ region of the green regions of Figs~\ref{fig:NP1}-\ref{fig:NP3}. A conclusive test the nature of the transition in the region of parameter space considered here may require a future $e^+e^-$ and/or $pp$ collider that is able to achieve a better than 5\% determination of $\Gamma(h\to\gamma\gamma)$ and a separate determination of $M_\Sigma$. One may also anticipate other loop-induced Higgs property deviations\footnote{We thank Lian-Tao Wang for raising this possibility.}, such as the rate for associated production $e^+e^-\to Z^\ast\to Zh$. 

Fig.~\ref{fig:NP3} contains contours of constant $T_c$ in the vicinity of the first order transition region. Knowledge of the critical temperature is interesting in its own right as well as for assessing the validity of the DR3EFT. We observe that for the parameter choices in the first order region, the physical triplet mass $M_\Sigma$ is greater than $T_c$, validating our treatment of the triplet as a heavy degree of freedom. Only for sufficiently large $M_\Sigma$ would the superheavy triplet DR3EFT be justified, giving {\it a posteriori} justification for concentrating on the heavy rather than superheavy case. 

Looking to the future, knowledge of $T_c$ will be important for assessing the strength of the phase transition in the light green region. We emphasize that our present study provides no information about the quantities that characterize the strength of the transition, such as the broken phase sphaleron rate relevant to electroweak baryogenesis or the latent heat and effective action relevant to the dynamics of gravitational radiation generated during a first order transition \cite{Caprini:2015zlo}. In principle, one could estimate the broken phase sphaleron rate using a combination of analytic and numerical methods (see Ref.~\cite{Patel:2011th} and references therein), a task that requires knowledge of the bubble nucleation temperature that is often reasonably approximated by $T_c$ but that goes beyond the scope of the present study. A more robust determination of the sphaleron rate would require a non-perturbative study. Similar comments apply to the thermodynamic quantities relevant to gravitational wave generation. We defer an in-depth analysis of these issues to future work.

\section{Discussion}
\label{sec:discuss}

In this paper, we have initiated a non-perturbative study of the electroweak phase transition in the $\Sigma$SM. 
We have performed a finite temperature dimensional reduction in this model, and derived a set of effective three-dimensional theories that can be studied by non-perturbative lattice simulations. We have immediately applied these effective theories in the case where triplet is assumed to be sufficiently heavy that it may be integrated out, leading to effective 3d theory of same form as in the SM, and existing lattice results of Ref.~\cite{Rummukainen:1998as} can be applied. We have found that there exist regions for which a one-step transition to the EWSB vacuum can be of first-order. In addition, for given value of triplet mass, there is a minimum value of the portal coupling that can accommodate a first-order transition.  Below this critical value the EWPT is a smooth crossover, as in the minimal SM.  We emphasize that in order to reach this conclusion, a non-perturbative treatment is crucial, since perturbative analyses cannot identify the existence of the crossover region. Furthermore, we have shown that the presence of a first order transition is associated with a lower bound on the $h\rightarrow\gamma\gamma$ partial width.  This bound would potentially allow one to probe regions of the parameter space allowing a first order EWPT with the high luminosity phase of the Large Hadron Collider or with a future $e^+ e^−$ and/or $pp$ collider.

We emphasize that our study of EWPT as it stands is limited to providing the critical temperature, and character of the EWPT (first order, second order, or crossover).  Without external information, certain thermodynamic properties, such as latent heat or bubble nucleation rate, relevant for the gravitational wave generation, or the broken phase sphaleron rate relevant to electroweak baryogenesis, can be inferred.

The existence of a crossover transition and the presence of a critical boundary between regions of crossover and first order transition can be revealed only in non-perturbative analysis.  Despite this, frequently used perturbative studies may potentially provide a reasonable qualitative, if not quantitative, agreement with lattice on other features of the EWPT.  In order to test the reliability of the perturbative approach, in part II we will perform a systematic comparative analysis of the performance of perturbation theory to extract thermodynamic quantities, which would allow us to set a definite benchmark for the accuracy of the perturbation theory.



\section*{Acknowledgments}
TT has been supported by the Vilho, Yrj\"{o} and Kalle
V\"{a}is\"{a}l\"{a} Foundation. LN and TT have been supported by the Academy of Finland grant no. 273545, as well as by the
European Research Council grant no. 725369. LN was also supported by the Academy
of Finland grant no. 308791. DJW (ORCID ID
0000-0001-6986-0517) was supported by Academy of Finland grant
no.~286769 and  and by the Research Funds of the University of Helsinki.  HHP and MJRM were supported in part under U.S. Department of Energy contract DE-SC0011095.  The authors would like to thank Keijo Kajantie, Jonathan Kozaczuk, Mikko Laine, Kari Rummukainen and Aleksi Vuorinen for discussions.

\appendix

\newpage
\section{Matching relations: $\Sigma$ heavy or light}
\label{sec:matching_light}

In this appendix, we list the matching relations of the normalization of fields and coupling constants between the three-dimensional effective theory at the heavy scale $\mathcal{L}_3$ in (\ref{lag_heavy}) and the full four-dimensional theory and the superheavy scale $\mathcal{L}$.  The relations are valid for case where the real triplet degrees of freedom $\Sigma^a$ are classified as either heavy or light.  Matching relations for the case where $\Sigma^a$ is superheavy are provided in the next section. 

In the following expressions, we use $N_d = 1$, $N_t = 1$ and $N_t = 3$ to identify contributions from the SM Higgs doublet, the real triplet $\Sigma$, and fermions.  Additionally, we make the following abbreviations arising from the evaluation of one and two loop integrals:
\begin{align}
L_b&=\ln\Big(\frac{\Lambda^2}{T^2}\Big)-2[\ln(4\pi)-\gamma], \\ 
L_f&= L_b+4\ln2, \\
c &= \frac{1}{2}\bigg(\ln\Big(\frac{8\pi}{9}\Big) + \frac{\zeta'(2)}{\zeta(2)} - 2 \gamma_E \bigg)
\end{align}

\subsection{Normalization of fields}

Here we collect normalizations between the four- and three-dimensional fields in the Landau gauge $\xi=0$. Field normalizations of $B_0$, $\vec{B}$ and $H$ are not affected by scalar triplet $\Sigma$, and are therefore same as in the SM. 
\begin{align}
\nonumber W_{\td,0}^2 &=\frac{W_{\fd,0}^2}{T}\bigg[1+\frac{g^2}{(4\pi)^2}\bigg(\frac{N_d+2 N_t-26}{6}L_b\\
&\hspace{5mm}+\frac{1}{3}(8+N_d+2 N_t)+\frac{4N_f}{3}(L_f-1)\bigg)\bigg], \\
\nonumber \vec{W}_{\td}^2 &=\frac{\vec{W}_{\fd}^2}{T}\bigg[1+\frac{g^2}{(4\pi)^2}\bigg(\frac{N_d + 2 N_t-26}{6}L_b\\
&\hspace{35mm}-\frac{2}{3}+\frac{4N_f}{3}L_f\bigg)\bigg], 
\end{align}

\begin{align}\label{eq:Bfieldtimenorm}
\nonumber B_{\td,0}^2
&=\frac{B_{\fd,0}^2}{T}\bigg[1+\frac{g'^2}{(4\pi)^2}\bigg(N_d\Big(\frac{L_b}{6}+\frac{1}{3}\Big)\\
&\hspace{35mm}+\frac{20N_f}{9}(L_f-1)\bigg)
\bigg],\\
\label{eq:Bfieldspacenorm}
\vec{B}_{\td}^2 &=\frac{\vec{B}_{\fd}^2}{T}\bigg[1+
  \frac{g'^2}{(4\pi)^2}\bigg(N_d\frac{L_b}{6}+\frac{20N_f}{9}L_f\bigg)\bigg].
\end{align}

\begin{align}\label{eq:higgsfieldnorm}
\nonumber \big(H^{\dagger}H\big)_\td &=\frac{\big(H^{\dagger}H\big)_\fd}{T}\bigg[1-\frac{1}{(4\pi)^2}\Big(\frac{3}{4}(3g^2 + {g'}^2)L_b \\
&\hspace{40mm}- 3 y^2_t L_f \Big)\bigg], \\
\label{eq:tripletfieldnorm}
\big(\Sigma^a \Sigma^a \big)_\td &=\frac{\big( \Sigma^a \Sigma^a \big)_\fd}{T}\bigg[1-\frac{1}{(4\pi)^2}\Big(6 g^2 L_b 
\Big)\bigg].
\end{align}

\subsection{Matching relations between superheavy and heavy scales}
\label{sec:matching_1loop}

Apart from Debye masses $m_D^2$, $m'^2_D$, $m''^2_D$, all parameters of the effective theory are calculated up to $O(g^4)$, which means one-loop accuracy for couplings and 2-loop accuracy for scalar mass parameters.  We have confirmed that to the order calculated, these relations are explicitly independent of the gauge parameter $\xi$.

The Debye masses for the SU(2), U(1), and SU(3) gauge fields, respectively are,
\begin{align}
m_D^2={}&g^2T^2\bigg(\frac{4+N_d+2N_t}{6}+\frac{N_f}{3}\bigg), \\
m'^2_D={}&g'^2T^2\bigg(\frac{N_d}{6}+\frac{5N_f}{9}\bigg),\\
m''^2_D={}&g^2_s T^2\bigg(1+\frac{N_f}{6}\bigg)\,.
\end{align}
Matching relations for the SU(2) and U(1) gauge coupling constants are
\begin{align}
\nonumber g_3^2={}&g^2(\Lambda)T\bigg[1 +\frac{g^2}{(4\pi)^2}\bigg(\frac{44-N_d -2 N_t}{6}L_b\\
&\hspace{25mm}+\frac{2}{3}-\frac{4N_f}{3}L_f\bigg)\bigg],\\
g'^2_3={}&g'^2(\Lambda)T\bigg[1 +\frac{g'^2}{(4\pi)^2}\bigg(-\frac{N_d}{6}L_b-\frac{20N_f}{9}L_f\bigg)\bigg]
\end{align}
The couplings among the temporal scalar fields are 
\begin{align}
\kappa_3={}&T\frac{g^4}{16 \pi^2} \frac{16+N_d +8 N_t-4N_f}{3}, \\
\kappa_3'={}&T\frac{g'^4}{16\pi^2} \bigg(\frac{N_d}{3}-\frac{380}{81} N_f\bigg),\\
\kappa_3''={}&T\frac{g^2g'^2}{16\pi^2}\bigg(2N_d-\frac{8}{3}N_f\bigg)\,.
\end{align}
\begin{widetext}
The couplings between temporal and fundamental/adjoint scalar fields are
\begin{align}
\notag
h_{3}={}&\frac{g^2(\Lambda)T}{4}\bigg(1+\frac{1}{(4\pi)^2}\bigg\{\bigg[\frac{44-N_d -2N_t}{6}L_b+\frac{53}{6}-\frac{N_d}{3} -\frac{2N_t}{3}-\frac{4N_f}{3}(L_f-1)\bigg]g^2+\frac{g'^2}{2} -6 y_t^2 + 12\lambda +8a_2 \bigg\} \bigg), \\
h'_{3}={}&\frac{g'^2(\Lambda)T}{4}\bigg(1 +\frac{1}{(4\pi)^2}\bigg\{\frac{3g^2}{2}+\bigg[\frac{1}{2}-\frac{N_d}{6}\Big(2+L_b \Big)  -\frac{20N_f}{9}(L_f-1)\bigg]g'^2 - \frac{34}{3} y_t^2 + 12\lambda \bigg\} \bigg), \\
h''_{3}={}&\frac{g(\Lambda)g'(\Lambda)T}{2}\bigg\{1+\frac{1}{(4\pi)^2}\bigg[-\frac{5+N_d}{6} g^2+ \frac{3-N_d}{6}g'^2+L_b\bigg(\frac{44-N_d}{12}g^2 -\frac{N_d}{12}g'^2\bigg)\notag \\
&-N_f(L_f-1)\bigg(\frac{2}{3}g^2+\frac{10}{9}g'^2\bigg) + 2 y_t^2 + 4 \lambda \bigg]\bigg\},\\
\omega_3={}&-\frac{2 T}{16 \pi^2} g^2_s y_t^2, \\
\delta_{3}={}& \frac{1}{2}g^2(\Lambda)T\bigg(1+\frac{1}{(4\pi)^2}\bigg\{ a_2 + 8 b_4 + g^2 \Big(\frac{16-N_d-2N_t}{3}-\frac{4}{3}N_f( L_f - 1) + L_b \frac{44-N_d-2N_t}{6} \Big) \bigg\} \bigg), \\
\delta_{3}'={}& -\frac{1}{2}g^2(\Lambda)T\bigg(1+\frac{1}{(4\pi)^2}\bigg\{ 4 b_4 + g^2 \Big(-\frac{20+N_d+2N_t}{3}-\frac{4}{3}N_f( L_f-1)+ L_b \frac{44-N_d-2N_t}{6} \Big) \bigg\} \bigg). \\
\end{align}

The matching relations for quartic couplings of the scalar potential are
\begin{align}
\lambda_3={}&T\Bigg\{\lambda(\Lambda) + \frac{1}{(4\pi)^2}\bigg[\frac{1}{8}\Big(3g^4 + {g'}^4 +2 g^2{g'}^2 \Big) + 3 L_f \Big(y_t^4 - 2\lambda y_t^2 \Big)  -L_b \bigg(\frac{3}{16}\Big(3g^4 + {g'}^4 + 2 g^2{g'}^2 \Big) \nonumber  \\
& - \frac{3}{2}\Big(3g^2+{g'}^2 -8 \lambda \Big) \lambda +\frac{3}{4}a^2_2 \bigg) \bigg]\Bigg\}, \\
\label{portal_heavy}a_{2,3}={}&T\Bigg\{a_2(\Lambda) + \frac{1}{(4\pi)^2}\bigg[2g^4  -3 a_2 y_t^2 L_f - L_b\Big(2a_2^2 + 5 a_2 b_4 + 3 g^4 + 6 a_2 \lambda - \frac{3}{4}a_2\big(g'^2 + 11 g^2\big)\Big)\bigg]\Bigg\}, \\
b_{4,3}={}&T\Bigg\{b_4(\Lambda) + \frac{1}{(4\pi)^2}\bigg[4g^4 - L_b\Big(a_2^2 + 11 b_4^2 - 12g^2 b_4 + 6 g^4 \Big)\bigg]\Bigg\}. \quad  
\end{align}
The matching relations for quartic couplings of the scalar potential are
\begin{multline}
\mu^2_3 = (\mu^2_3)_\text{SM} + \frac{T^2}{8}a_2(\Lambda)
+ \frac{1}{16\pi^2} \bigg\{ +\frac{3}{2} a_2 \mu^2_{\Sigma} L_b  + T^2 \bigg( \frac{5}{24}g^4 + \frac{1}{2} a_2 g^2 - \frac{3}{8}a_2 y_t^2 L_f + L_b \Big( -\frac{7}{16}g^4 - \frac{5}{8}a^2_2 \\
 - \frac{5}{8}a_2 b_4 + \frac{33}{32} a_2 g^2 + \frac{3}{32}a_2{g'}^2 - \frac{3}{4}a_2 \lambda \Big)  + \Big( c + \ln(\frac{3T}{\Lambda_{3d}}) \Big)\Big(-\frac{3}{2}a^2_{2,3} + 6 a_{2,3} g_3^2  -\frac{3}{4}g_3^4   
\Big) \bigg) \bigg\} ,
\end{multline}
where (see \cite{Gorda:2018hvi, Kajantie:1995dw})
\begin{multline}
(\mu^2_3)_\text{SM} = -\mu^2(\Lambda) +\frac{T^2}{16}\Big(3g^2(\Lambda) + {g'}^2(\Lambda) + 4 y_t^2(\Lambda) + 8 \lambda(\Lambda) \Big) + \frac{1}{16\pi^2} \bigg\{-\mu^2\bigg( \Big(\frac{3}{4}(3g^2 + {g'}^2) - 6 \lambda \Big)L_b - 3 y_t^2 L_f \bigg) \\
 + T^2 \bigg( \frac{167}{96}g^4 + \frac{1}{288}{g'}^4 - \frac{3}{16}g^2{g'}^2 + \frac{1}{4}\lambda(3g^2+{g'}^2)
 + L_b \Big( \frac{17}{16}g^4 - \frac{5}{48}{g'}^4 - \frac{3}{16}g^2{g'}^2 + \frac{3}{4}\lambda(3g^2+{g'}^2) - 6 \lambda^2 \Big) \\
 + \Big( c + \ln(\frac{3T}{\Lambda_{3d}}) \Big)\Big( \frac{39 g_3^4}{16} -\frac{5g_3'^4}{16}-\frac{9}{8} g_3^2 g_3'^2 + 12 g_3^2 h_3 - 6 h_3^2 - 2 h_3'^2-3 h_3''^2
 + 3 \lambda_3(3g_3^2+g_3'^2) - 12 \lambda_3^2
\Big) \\
 - y_t^2 \Big(\frac{3}{16}g^2 + \frac{11}{48}{g'}^2 + 2 g^2_s \Big) + (\frac{1}{12}g^4 + \frac{5}{108}{g'}^4)N_f 
 + L_f \Big( y_t^2 \Big(\frac{9}{16}g^2 + \frac{17}{48}{g'}^2 + 2 g^2_s - 3 \lambda \Big) +\frac{3}{8}y_t^4 - (\frac{1}{4}g^4 + \frac{5}{36}{g'}^4 ) N_f \Big) \\
 + \ln(2) \Big( y_t^2 \Big(-\frac{21}{8}g^2 - \frac{47}{72}{g'}^2 + \frac{8}{3} g^2_s + 9 \lambda \Big) -\frac{3}{2}y_t^4 + (\frac{3}{2}g^4 + \frac{5}{6}{g'}^4 ) N_f \Big) \bigg) \bigg\}\,,
\end{multline}
and
\begin{multline}
\mu^2_{\Sigma,3} = -\mu^2_{\Sigma} + T^2 \Big( \frac{1}{6}a_2(\Lambda) + \frac{5}{12}b_4(\Lambda) + \frac{1}{2}g^2(\Lambda) \Big)
 \frac{1}{16\pi^2} \bigg\{- \Big(6 g^2 - 5 b_4 \Big) \mu^2_{\Sigma} L_b + 2 \mu^2 a_2 L_b \\
 + T^2 \bigg( \Big(\frac{71}{18} + \frac{2}{9} N_f\Big) g^4 + \frac{5}{3} b_4 g^2 + \frac{1}{4} a_2 {g}^2 +\frac{1}{12} a_2 {g'}^2
 + L_b \Big( \frac{5}{12}g^4 - \frac{3}{4}a^2_2 - \frac{55}{12}b^2_4 + \frac{11}{8} a_2 g^2 + \frac{1}{8}a_2{g'}^2 + 5 b_4 g^2 - \frac{5}{6}a_2 b_4 - a_2 \lambda \Big) \\
 + \Big( c + \ln(\frac{3T}{\Lambda_{3d}}) \Big)\Big( -2 a^2_{2,3}-10 b^2_{4,3}+a_{2,3}(3g_3^2+g_3'^2)+20b_{4,3} g_3^2-3g_3^4
+24g_3^2 \delta_3-24\delta_3^2+8g_3^2\delta_3'-16\delta_3 \delta_3'-16\delta_3'^2
\Big) \\
 - L_f \Big(\frac{1}{2}a_2 y_t^2 + \frac{2}{3}g^4 N_f \Big) + \ln(2) \Big(3a_2 y_t^2 + 4 g^4 N_f \Big)\bigg) \bigg\}\,.
\end{multline}

\subsection{Matching relations between heavy and light scales}
\label{sec:matching_light}

Below we list matching relations for final 3d theory parameters, when heavy time components of the gauge fields $B_0$,$W_0^a$ and $G_0^A$ are integrated out, assuming that both the Higgs doublet and triplet mass parameters are light. 

\begin{align}
\bar{g}^2_3 =& g^2_3 \Big( 1 - \frac{g^2_3}{24\pi m_D} \Big), \\
\bar{g}'^2_3 =& g'^2_3, \\
\bar{\lambda}_{3} =& \lambda_{3} - \frac{1}{8\pi}\Big( \frac{3 h^2_{3}}{m_D} + \frac{ h_{3}'^2 }{m_D'} + \frac{ h_{3}''^2}{m_D+m_D'} \Big), \\
\bar{a}_{2,3} =& a_{2,3} - \frac{h_3}{2 \pi m_D}(3\delta_3 + \delta_3') ,\\ 
\bar{b}_{4,3} =& b_{4,3} -\frac{1}{2 \pi m_D}(3\delta_3^2 + 2 \delta_3 \delta_3' + {\delta_3'}^2) ,\\ 
\bar{\mu}^2_{3} =& \mu^2_3  -\frac{1}{4\pi}\Big(3 h_{3} m_D +  h_{3}' m_D' + 8 \omega_3 m_D'' \Big) + \frac{1}{16\pi^2} \bigg( 3g^2_3 h_{3} - 3 h^2_{3} - {h_3'}^2 - \frac{3}{2}{h_3''}^2 + 2\mu_{3}\Big( 3\frac{{h_3}^2}{m_D}  +  \frac{{h_3'}^2}{m_D'} \Big) \nonumber \\
& \quad + \Big(-\frac{3}{4}g^4_3 + 12 g^2_3 h_{3} \Big) \ln\Big(\frac{\Lambda_{3d}}{2m_D} \Big)  -  6 h^2_{3} \ln\Big(\frac{\Lambda_{3d}}{2m_D+\mu_{3}} \Big) - 2 {h_3'}^2 \ln\Big(\frac{\Lambda_{3d}}{2m_D'+\mu_{3}} \Big) \nonumber \\
& \quad - 3 {h_3''}^2 \ln\Big(\frac{\Lambda_{3d}}{m_D+m_D'+\mu_{3}} \Big) \bigg) + \frac{3}{16\pi^2} h_3 (3\delta_3 + \delta_3') \frac{\mu_{\Sigma,3}}{m_D}, \\ 
\bar{\mu}^2_{\Sigma,3} =& \mu^2_{\Sigma,3}+  \frac{m_D}{2\pi}(3\delta_3 + \delta_3') + \frac{1}{16\pi^2} \bigg( 2g^2_3(3\delta_3+\delta_3') - 12 \delta^2_3 - 12 \delta_3 \delta'_3 - 8 {\delta_3'}^2  + 4 h_3 (3\delta_3 + \delta_3') \frac{\mu_{3}}{m_D} \nonumber \\
& \quad + ( 18 \delta^2_3 + 12 \delta_3 \delta'_3 + 2 {\delta_3'}^2 )\frac{\mu_{\Sigma,3}}{m_D} + \Big(-2g^4_3 + 8 g^2_3(3\delta_3 + \delta_3')  \Big) \ln\Big(\frac{\Lambda_{3d}}{2m_D} \Big)\nonumber \\
& \quad  -(24 \delta^2_3 + 24 \delta_3 \delta'_3 + 16 {\delta_3'}^2 ) \ln\Big(\frac{\Lambda_{3d}}{2m_D+\mu_{\Sigma,3}} \Big) \bigg).
\end{align}

\subsection{Matching relations between heavy and light scales where heavy triplet is integrated out}
\label{sec:heavy_triplet}

Below we list matching relations for the light scale DR3EFT  parameters, where the zero Matsubara mode of the real triplet $\Sigma$ is integrated out simultaneously with the time components of the gauge fields $B_0$, $W_0^a$, $G_0^A$.
\begin{align}
\bar{g}^2_3 =& g^2_3 \Big( 1 - \frac{g^2_3}{24\pi}\Big(\frac{1}{ \mu_{\Sigma,3}} + \frac{1}{m_D}\Big)\Big), \\
\bar{g}'^2_3 =& g'^2_3, \\
\bar{\lambda}_{3} =& \lambda_{3} - \frac{1}{8\pi}\Big( \frac{3 h^2_{3}}{m_D} + \frac{ h_{3}'^2 }{m_D'} + \frac{ h_{3}''^2}{m_D+m_D'} + \frac{3 a^2_{2,3}}{4 \mu_{\Sigma,3}}\Big), 
\end{align}
\begin{align}
\bar{\mu}^2_{3} =& \mu^2_3 -\frac{1}{4\pi}\Big(3 h_{3} m_D +  h_{3}' m_D' + \frac{3a_{2,3} \mu_{\Sigma,3}}{2} \Big) \nonumber \\
&+\frac{1}{16\pi^2}\bigg\{\Big(3g^2_3 h_{3} - 3 h^2_{3} - {h_3'}^2 - \frac{3}{2}{h_3''}^2 + 2\mu_{3}\Big( 3\frac{{h_3}^2}{m_D}  +  \frac{{h_3'}^2}{m_D'} \Big) \nonumber \\
& \quad + \Big(-\frac{3}{4}g^4_3 + 12 g^2_3 h_{3} \Big) \ln\Big(\frac{\Lambda_{3d}}{2m_D} \Big)  -  6 h^2_{3} \ln\Big(\frac{\Lambda_{3d}}{2m_D+\mu_{3}} \Big) \nonumber \\
& \quad - 2 {h_3'}^2 \ln\Big(\frac{\Lambda_{3d}}{2m_D'+\mu_{3}} \Big) - 3 {h_3''}^2 \ln\Big(\frac{\Lambda_{3d}}{m_D+m_D'+\mu_{3}} \Big)\bigg\}_{\text{SM}} \nonumber \\
&+\frac{1}{16\pi^2} \bigg\{\Big(\frac{3h_3\mu^2_{\Sigma,3}}{m_D} + \frac{3 a_{2,3}m_D}{2\mu^2_{\Sigma,3}}\Big)(3\delta_3+\delta_3') - \frac{3}{4} a^2_{2,3}+ \frac{15}{4} a^2_{2,3} b_{4,3} \nonumber \\
&\quad+ \frac{3}{2} a_{2,3} g^2_{3}+ \frac{3}{2} a^2_{2,3} \frac{\mu_{3}}{\mu_{\Sigma,3}} + \Big(6 a_{2,3} g^2_3 - \frac{3}{4}g^4_3 \Big) \ln\Big(\frac{\Lambda_{3d}}{2\mu_{\Sigma,3}}\Big) -\frac{3}{2} a^2_{2,3} \ln\Big(\frac{\Lambda_{3d}}{2\mu_{\Sigma,3}+\mu_{3}}\Big)
\bigg\}_{\text{$\Sigma$SM}}
\end{align}

\section{Matching relations in the case of superheavy triplet}
\label{sec:matching_superheavy}
In the case that the mass parameter of the real triplet is large and negative, $|\mu^2_\Sigma|\gtrsim (\pi T)^2$, the real triplet degrees of freedom are classified as superheavy, and all their Matsubara modes (including the zero mode) are integrated out to derive the heavy scale DR3EFT.  Matching relations for parameters of the resulting 3d theory require the evaluation sum-integrals involving the real triplet.  Because the two-loop sum integrals are technically difficult, we have carried out the matching to only $\mathcal{O}(g^2)$ for the mass parameters $\mu_3^2$ and $\mu_{\Sigma,3}^2$.  The relations below are written in terms of derivatives of the bosonic thermal function
\begin{equation}
J_B(z^2) = \int_0^\infty dx\, x^2 \ln (1-e^{-\sqrt{x^2 + z^2}})
\end{equation}
evaluated at $z^2 = |\mu_\Sigma^2|/T^2$.

The normalizations of the SU(2) gauge fields are:
\begin{align}
W_{\td,0}^2 &=\frac{W_{\fd,0}^2}{T}\bigg[1+\frac{g^2}{(4\pi)^2}\bigg(\frac{N_d-26}{6}L_b+\frac{1}{3}(8+N_d)+\frac{4N_f}{3}(L_f-1) \nonumber \\
& \quad + \frac{N_t}{3} \Big( 16\pi^2(\frac{-1}{\pi^2}J_B'' + \frac{4 \mu^2_\Sigma}{2\pi^2 T^2} J_B''' +  \ln\Big|\frac{\Lambda^2}{\mu^2_\Sigma}\Big| \Big) \bigg)\bigg], \\
\vec{W}_{\td}^2 &=\frac{\vec{W}_{\fd}^2}{T}\bigg[1+\frac{g^2}{(4\pi)^2}\bigg(\frac{N_d -26}{6}L_b-\frac{2}{3}+\frac{4N_f}{3}L_f + \frac{N_t}{3} \Big(-16 J_B'' + \ln\Big|\frac{\Lambda^2}{\mu^2_\Sigma}\Big| \Big) \bigg)\bigg], 
\end{align}
Normalizations of all other fields do not depend on the real triplet, and are therefore identical to those listed in (\ref{eq:Bfieldtimenorm})--(\ref{eq:higgsfieldnorm}).

The parameters of the heavy scale DR3EFT which are modified by superheavy triplet are listed below.   And other relations remain same as in earlier section.
\begin{align}
m_D^2&=g^2T^2\bigg(\frac{4+N_d}{6}+\frac{N_f}{3} + \frac{4N_t}{\pi^2} \Big( J'_B + \frac{\mu^2_\Sigma}{T^2} J''_B \Big)\bigg),  \\
g_3^2&=g^2(\Lambda)T\bigg[1 +\frac{g^2}{(4\pi)^2}\bigg(\frac{44-N_d}{6}L_b+\frac{2}{3}-\frac{4N_f}{3}L_f -\frac{N_t}{3}\Big(-16 J''_B + \ln\Big|\frac{\Lambda^2}{\mu^2_\Sigma}\Big| \Big)\bigg)\bigg],\\
\kappa_3&=T g^4 \bigg[\frac{1}{16 \pi^2} \bigg(\frac{16+N_d -4N_f}{3}\bigg) + \frac{8 (\mu^2_\Sigma)^2}{3\pi^2 T^4} J_B'''' \bigg], \\
\notag
h_{3}&=\frac{g^2(\Lambda)T}{4}\bigg(1+\frac{1}{(4\pi)^2}\bigg\{\bigg[\frac{44-N_d}{6}L_b+\frac{53}{6}-\frac{N_d}{3}-\frac{4N_f}{3}(L_f-1) \nonumber \\
& \quad - \frac{N_t}{3}\Big( -16(J_B'' +\frac{2 \mu^2_\Sigma}{\pi^2 T^2} J_B''') + \ln\Big|\frac{\Lambda^2}{\mu^2_\Sigma}\Big| \Big) \bigg]g^2+\frac{g'^2}{2} -6 y_t^2 + 12\lambda + 128 a_2 \frac{\mu^2_\Sigma}{T^2} J_B''' \bigg\} \bigg), \\
\mu^2_3&=-\mu^2(\Lambda)+T^2\bigg(\frac{1}{16}(3g^2(\Lambda)+g'^2(\Lambda))+\frac{1}{4} y_t^2(\Lambda) + \frac{1}{2} \lambda(\Lambda) \bigg) + \frac{3}{2}a_2(\Lambda) \bigg(\frac{T^2}{\pi^2}J_B' + \frac{\mu^2_\Sigma}{16\pi^2}\Big(1+\ln\Big|\frac{\Lambda^2}{\mu^2_\Sigma}\Big| \Big) \bigg), \\
\lambda_3&=T\Bigg\{\lambda(\Lambda) + \frac{1}{(4\pi)^2}\bigg[\frac{1}{8}\Big(3g^4 + {g'}^4 +2 g^2{g'}^2 \Big) + 3 L_f \Big(y_t^4 - 2\lambda y_t^2 \Big) \nonumber  \\
& -L_b \bigg(\frac{3}{16}\Big(3g^4 + {g'}^4 + 2 g^2{g'}^2 \Big) - \frac{3}{2}\Big(3g^2+{g'}^2 -8 \lambda \Big) \lambda\bigg) -\frac{3}{4}a^2_2 \Big(-16 J_B' + \ln\Big|\frac{\Lambda^2}{\mu^2_\Sigma}\Big| \Big)  \bigg]\Bigg\}.
\end{align}
Matching relations for the parameters of the light scale DR3EFT for the superheavy triplet remain the same as in Appendix \ref{sec:heavy_triplet} above.

\begin{figure}
\centering
\includegraphics[width=0.6\textwidth]{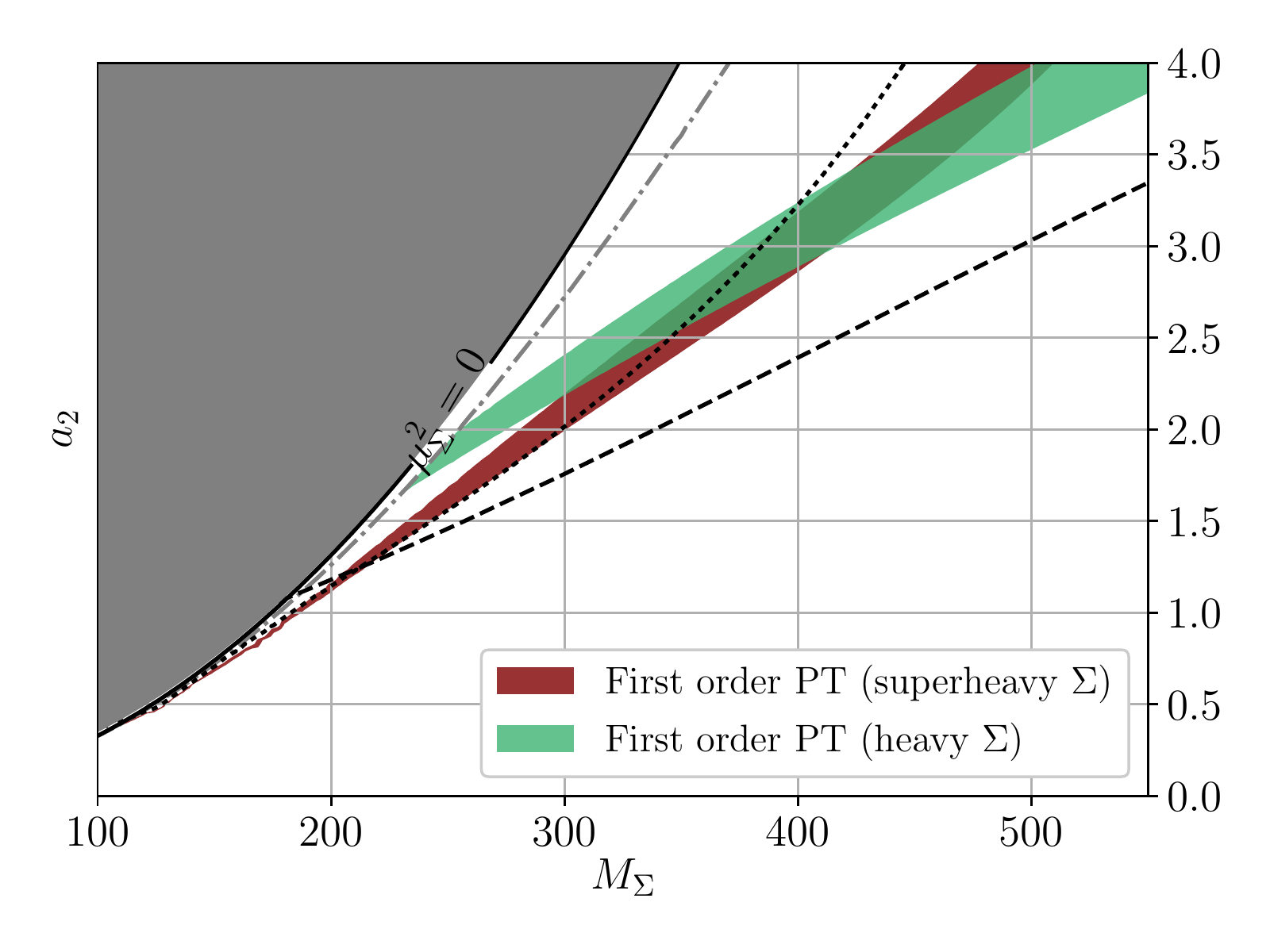}
\caption{Comparison of heavy and superheavy $\Sigma$ approximations. Gray dot-dashed line shows loop corrected $\mu^2_\Sigma = 0$ curve, and black dashed and dotted curves show $T_c = 140$ GeV for heavy and superheavy cases, respectively.}
\label{fig:superheavy}
\end{figure}

In Fig.~\ref{fig:superheavy} we show a comparison between heavy and superheavy approximations. First order transition region is again given by 3d parameter $0<x<0.11$. Black dashed and dotted curves show $T_c = 140$ GeV for heavy and superheavy cases, respectively.  We observe that locations of first order regions agree qualitatively, while $T_c$ curves show larger discrepancy. We assume that this difference in critical temperatures is related to our approximation in the superheavy case, where we only used one-loop level determination for mass parameter that gives $y$, from which $T_c$ is solved.

\section{Counterterms of the 3d effective theories}
\label{app:counterterms}
In this section, we collect the counterterms associated with the logarithmic UV divergences of the 3d effective theory.
The UV-divergent parts can be extracted by a direct diagrammatic calculation of the scalar self energies at zero external momentum at two-loop. 
At the DR3EFT at the heavy scale, the mass parameter counterterm for the doublet is
\begin{align}
\delta \mu^2_3 = (\delta \mu^{2}_{3})_\text{SM} - \frac{1}{16\pi^2}\frac{1}{4\epsilon} \Big(-\frac{3}{4}g^4_3 + 6 a_{2,3} g^2_3 - \frac{3}{2}a^2_{2,3} \Big),
\end{align}
where the pure Standard Model contribution is 
\begin{align}
(\delta \mu^{2}_{3})_\text{SM} =& -\frac{1}{16\pi^2}\frac{1}{4\epsilon} \bigg( \frac{39}{16}g^4_3 + 12g^2_3 h_{3} - 6 h^2_{3} + 9 g^2_3 \lambda_{3} - 12\lambda^2_{3} -\frac{5}{16}{g_3'}^4 - \frac{9}{8}g^2_3 {g_3'}^2 - 2{h_3'}^2 - 3{h_3''}^2 + 3 {g'}^2_3 \lambda_{3} \bigg),
\end{align}
and the mass parameter counterterm for the real triplet is 
\begin{multline}
\delta \mu^2_{\Sigma,3} = -\frac{1}{16\pi^2}\frac{1}{4\epsilon} \bigg(-3g^4_3 + 8 g^2_3(3\delta_3 + \delta_3') -8(3\delta^2_3 + 2 \delta_3 \delta_3' + 2{\delta_3'}^2) 
+ a_{2,3}(3g^2_3 + {g'}^2_3) + 20 b_{4,3}g^2_3 - 2 a^2_{2,3} - 10 b^2_{4,3} \bigg).
\end{multline}

In the DR3EFT at the light scale, the mass parameter counterterm for the doublet is
\begin{align}
\delta \bar{\mu}^2_{3}= \delta \bar{\mu}^{2,\text{SM}}_{3} - \frac{1}{16\pi^2}\frac{1}{4\epsilon} \Big(-\frac{3}{4}\bar{g}^4_3 + 6 \bar{a}_{2,3} \bar{g}^2_3 - \frac{3}{2}\bar{a}^2_{2,3} \Big),
\end{align}
where the Standard Model contribution is
\begin{align}
\delta \bar{\mu}^{2,\text{SM}}_{3} = -\frac{1}{16\pi^2}\frac{1}{4\epsilon} \bigg( \frac{51}{16}\bar{g}^4_3+ 9 \bar{g}^2_3 \bar{\lambda}_{3} - 12\bar{\lambda}^2_{3} -\frac{5}{16}{\bar{g_3'}}^4 - \frac{9}{8}\bar{g}^2_3 {\bar{g_3'}}^2 + 3 {\bar{g_3'}}^2 \bar{\lambda}_{3} \bigg),
\end{align}
and the mass parameter counterterm for the real triplet is 
\begin{align}
\delta \bar{\mu}^2_{\Sigma,3} =& -\frac{1}{16\pi^2}\frac{1}{4\epsilon} \bigg(-\bar{g}^4_3 + \bar{a}_{2,3}(3\bar{g}^2_3 + {\bar{g'}}^2_3) + 20 \bar{b}_{4,3}\bar{g}^2_3 - 2 \bar{a}^2_{2,3} - 10 \bar{b}^2_{4,3} \bigg).
\end{align}

\end{widetext}

\bibliography{TripletEWPTRefs}

\end{document}